\documentclass[manuscript]{aastex6}
\usepackage{bm}

\defcitealias{Edmondson2010b}{Paper~I}

\shorttitle{3D Reconnection with a Guide Field}
\shortauthors{Edmondson \& Lynch}

\begin{document}


\title{Formation and Reconnection of Three-dimensional Current Sheets with a Guide Field in the Solar Corona}


\author{J. K. Edmondson\altaffilmark{1} and B. J. Lynch\altaffilmark{2}}







\altaffiltext{1}{Department of Climate and Space Sciences and Engineering, University of Michigan, 2455 Hayward Avenue, Ann Arbor, MI 48109, USA}
\altaffiltext{2}{Space Sciences Laboratory, University of California--Berkeley, 7 Gauss Way, Berkeley, CA 94720, USA}

\begin{abstract}

We analyze a series of three-dimensional magnetohydrodynamic numerical 
simulations of magnetic reconnection in a model solar corona to study the 
effect of the guide field component on quasi-steady state interchange 
reconnection in a pseudostreamer arcade configuration. 
This work extends the analysis of \citet{Edmondson2010b} by quantifying 
the mass density enhancement coherency scale in the current sheet associated 
with magnetic island formation during the nonlinear phase of plasmoid-unstable 
reconnection. 
We compare the results of four simulations of a zero, weak, moderate, and a strong guide field, 
$B_{GF}/B_0 = \lbrace 0.0 , 0.1 , 0.5 , 1.0 \rbrace$, to quantify the plasmoid density enhancement's 
longitudinal and transverse coherency scales as a function of the guide field strength.
We derive these coherency scales from autocorrelation and wavelet analyses, and 
demonstrate how these scales may be used to interpret the density enhancement fluctuation's 
Fourier power spectra in terms of a structure formation range, an energy 
continuation range, and an inertial range---each population with a distinct spectral slope.
We discuss the simulation results in the context of solar and heliospheric 
observations of pseudostreamer solar wind outflow and possible signatures of 
reconnection-generated structure.
  

\end{abstract}




\section{Introduction} \label{sec:intro}

Magnetic reconnection is one of the most interesting and fundamental plasma 
physics processes of mass, momentum, and energy transfer observed in geophysical, 
astrophysical, and laboratory plasmas 
\citep[see reviews by][and references therein]{Zweibel2009, Pontin2011, Daughton2012}. 
Since the introduction of the fractal reconnection interpretation by 
\citet{Shibata2001} of the well-known tearing-mode \citep{Furth1963,Biskamp1986}, 
there has been significant advances in theoretical and numerical simulation work examining the 
dynamics and evolution of resistive tearing and magnetic island creation, now 
often referred to as the plasmoid instability, including the formation and subsequent 
fragmentation of secondary current sheets during the system's transition to fully 
nonlinear evolution 
\citep[e.g.][]{Loureiro2007, Loureiro2012, Lapenta2008, Huang2010, Uzdensky2010, Barta2011, Murphy2013, PucciVelli2014, Teneranietal2016}.
In three-dimensions (3D), the additional degree of freedom allows for a much broader 
range of dynamics available to the system undergoing reconnection---everything from the 
interaction of magnetic flux ropes \citep{Linton2001,Linton2003}, the role of the guide 
field on particle and field evolution in and around the diffusion region 
\citep{Pritchett2004, Karimabadi2005, Drake2006, Drake2014}, to the macroscopic 
evolution of 3D tearing and reconnection along the current sheet 
\citep[e.g.][]{Linton2009, Shimizu2009, Shepherd2012, Cassak2013}, and the development of turbulence as a result of the reconnection itself \citep[see review by][and references therein]{KarimabadiLazarian2013}.

\citet{Nishida2013} examined the formation of 3D magnetic island plasmoids in an eruptive 
flare current sheet and \citeauthor{Wyper2014b} (\citeyear{Wyper2014b,Wyper2014a}) 
have shown the evolution of magnetic island flux rope structures can result in 
exceedingly complex dynamics, even when the 3D islands are more-or-less confined 
to a quasi-2D surface or within adjacent boundary layers.
Additionally, \citet{Baalrud2012} have shown that in 3D, the unstable modes of the plasmoid 
instability can grow at oblique angles which can interact and lead to a kind of 
``self-generated'' turbulent reconnection 
\citep[e.g.][]{Oishietal2015ApJ,HuangBhattacharjee2016ApJ} and \citet{Wangetal2015ApJ} 
have shown that the efficiency of magnetic energy conversion in 3D MHD reconnection 
is increased by similar nonlinear coupling between the inflow and outflow regions of multiple 
tearing layers that form in a global current sheet.

\citet{Edmondson2010b} showed the self-consistent formation and evolution of (very short) three-dimensional plasmoidal flux ropes within a singular reconnection current sheet in MHD. While this system exhibited turbulent characteristics, the effect of the plasmoids on the evolution of the system, however, could not be addressed. Several kinetic studies of three-dimensional reconnection layers in anti-parallel and finite guide field geometries have shown the evolution is always dominated by the formation and interaction of secondary islands \citep{Daughtonetal2011a,Daughtonetal2011b}. The flux ropes are produced by secondary instabilities within the electron layers, occur in both collisional regimes and large-scale collisionless systems, and lead to turbulent evolution. In particular, in the collisional regime the secondary islands break the Sweet-Parker scaling rate driving the system to kinetically-dominated scales.

Recently, \citet{Fujimoto2016} highlighted the intermittent, quasi-turbulent formation and ejection of 
3D magnetic island plasmoid flux ropes in a particle-in-cell simulation, obtaining results 
that are qualitatively similar to the MHD simulation results we present here. 
Analysis of magnetic reconnection processes at MHD scales is valuable specifically 
because of the intrinsic coupling of disparate spatial scales associated with the diffusion region 
and the macroscopic evolution of global system 
in which the both large-scale magnetic field and bulk plasma dynamics directly 
influence the reconnection process.
The most obvious example of this intrinsic scale-coupling in the corona is in solar flares 
\citep[e.g. see][and references therein]{Janvier2017}, however, there is mounting evidence that magnetic 
reconnection processes are also likely to be important to the origin of the slow solar wind.  

The basic, steady-state picture of slow solar wind generation from coronal helmet streamers 
and pseudostreamers has been established 
\citep[e.g.][]{Sheeley1999, WangYM2000, WangYM2012, Riley2012}. The static (i.e., potential) 
magnetic field structure of coronal pseudostreamers is expected to produce a stationary slow wind 
due to the flux tube expansion factors \citep{Arge2000, Cranmer2012, WangYM2012}. 
A prediction of the Separatrix-Web model \citep{Antiochos2011, Antiochos2012, Titov2012} 
is that pseudostreamer slow wind forms along arcs that can extend large angular distances 
away from the main helmet streamer belt and recent in situ data analysis appear to confirm this 
scenario \citep{Crooker2012a, Crooker2014, Owens2014}. 

In situ observations of slow solar wind plasma, field, and composition and their 
variability---including during periods of pseudostreamer wind---are strongly 
suggestive of a closed-field coronal source region which implies some magnetic 
reconnection process as a crucial component of its origin 
\citep{Zurbuchen2000,Zurbuchen2002,Kepko2016}. 
\citet{Higginson2017c} have demonstrated that the 3D structure of reconnection-generated magnetic islands in the heliospheric current sheet are consistent with the observations of small-scale flux rope transients in the slow solar wind \citep[][]{Feng2008, Cartwright2008, Cartwright2010, Kilpua2009b, Yu2014}. And \citet{Janvier2014} have argued that the power-law distribution of the sizes of these small flux rope transients appears compatible with the power-law distributions of magnetic island plasmoids obtained in various reconnection studies \citep[e.g.][]{Uzdensky2010,Fermo2010,Huang2012,Loureiro2012,Shen2013a,Lynch2016a}.

The magnetic topology of a pseudostreamer is a region of opposite polarity embedded in unipolar field \citep[e.g.,][]{Edmondson2009,Edmondson2010a,Lynch2016a}. The bipolar, 
closed-flux system has a magnetic null point and separatrix dome defining the boundary 
between the closed and open flux systems at low-to-moderate heights in the atmosphere 
\citep{WangYM2007a, WangYM2012, Titov2012, Rachmeler2014}.
We, and many other researchers, have shown that this pseudostreamer topology is ideally 
suited for the formation of current sheets and magnetic reconnection in response to global 
stresses accumulated via photospheric motions or other gradual energization processes 
\citep[e.g., see][and references therein]{Masson2014b}.
Thus, interchange reconnection as a dynamic and transient release mechanism of 
previously closed-flux plasma into the open field continues to be explored in remote 
sensing observations and in numerical simulations 
\citep{Baker2009, Edmondson2010a, vanDrielGesztelyi2012, Rappazzo2012, Masson2012, Masson2014b, Lynch2014, Pontin2015, Viall2015, Higginson2017a, Higginson2017b}. 

Here we present an extension of the \citet{Edmondson2010b} investigation, hereafter 
referred to as \citetalias{Edmondson2010b}, by performing a set of new high-resolution, 
adaptively-refined 3D MHD simulations with a systematically increasing guide field 
component to examine the role of the guide field on the development and evolution 
of the reconnection.
Our simulations start from a true 3D magnetic null line (X-type null point in 2D cross-section) 
and follow the formation and evolution of the current sheet in response to global driving 
far from the magnetic null line. Thus, our current sheet formation and reconnection 
are a self-consistent response to the imposed separation of the pseudostreamer arcade spine lines (fan surfaces) in the 
Syrovatskii scenario 
\citep{Syrovatskii1978a,Syrovatskii1978b,Syrovatskii1981,Antiochos2002}.

%
%
The paper is organized as follows. In section~\ref{S:CompDet}, we describe the 
numerical methods, the magnetic configurations, and boundary conditions for our 
set of MHD simulations. In section~\ref{S:Results}, we present our simulation 
results including the transition to the non-linear phase of the plasma instability 
under ``steady state'' reconnection ($\S$~\ref{SS:NonLinOnset}), the effects of 
the guide field on the characteristics of 3D plasmoid magnetic island flux ropes 
($\S$~\ref{SS:FluxRopeForm}), and the quantitative comparison of the coherency 
scales of the reconnection-generated plasmoid densities in each of the current 
sheets ($\S$~\ref{SS:CharCoherScl}). In section~\ref{S:Discussion}, we discuss 
our results in terms of reconnection-generated solar wind outflow at psuedostreamers 
and in section~\ref{S:Conc}, we present our conclusions. 

\section{Numerical Simulations} \label{S:CompDet}

We use the Adaptively Refined MHD Solver \citep[ARMS;][]{DeVore2008} to solve the ideal MHD equations in Cartesian coordinates,
\begin{equation}
\frac{\partial \rho}{\partial t} + \nabla \cdot \left( \ \rho \ \bm{V} \ \right) = 0 \ ,
\end{equation}
\begin{equation}
\frac{\partial \left( \ \rho \ \bm{V} \ \right)}{\partial t} + \nabla \cdot \left( \ \rho \ \bm{V} \ \bm{V} \ \right) + \nabla P = \frac{1}{4 \pi} \ \left( \ \nabla \times \bm{B} \ \right) \times \bm{B} \ ,
\end{equation}
\begin{equation}
\frac{\partial T}{\partial t} + \nabla \cdot \left( \ T \ \bm{V} \ \right) + \left( \ 1 - \gamma \ \right) \ T \ \left( \ \nabla \cdot \bm{V} \ \right) = 0 \ ,
\end{equation}
\begin{equation}
\frac{\partial \bm{B}}{\partial t} = \nabla \times \left( \ \bm{V} \times \bm{B} \ \right) \ .
\end{equation}
The variables retain their usual meaning: mass density $\rho$, velocity vector 
$\bm{V}$, magnetic field vector $\bm{B}$, and we have written the energy 
equation in terms of the plasma temperature $T$. The ratio of specific heats is 
$\gamma = 5/3$. The system is closed with the ideal gas law,
\begin{equation} \label{E:IdealGas}
P = 2 \left( \frac{\rho}{m_{p}} \right) k_{B} T \ ,
\end{equation}
where $m_{p}$ is the mass of a proton and $k_{B}$ is the Boltzmann constant.

ARMS utilizes block-adaptive computational gridding based on the PARAMESH toolkit \citep{MacNeice2000} for static and dynamic grid refinement and is optimized for modern massively parallel supercomputing architectures. ARMS has been used to study numerous dynamic phenomena in the solar atmosphere, including coronal jets \citep{Pariat2015, Wyper2016b, Karpen2017}, theinteraction between closed and open fields at streamer boundaries \citep{Edmondson2009, Edmondson2010a, Masson2013, Higginson2017a, Higginson2017b}, and the examination of plasmoid-unstable reconnection and magnetic island evolution (\citetalias{Edmondson2010b}; \citealt{Guidoni2016}; \citealt{Lynch2016a}).
 
\subsection{Initial Configuration, Driving Flows, and Boundary Conditions}
\label{SS:IBC}

The initial model atmosphere consists of a uniform mass density, $\rho_{0} = 1.6726 \times 10^{-16}$~g~cm$^{-3}$, and pressure, $P_{0} = 0.035$~dyne~cm$^{-2}$, that result in an initial uniform temperature of $T_0 = 1.27 \times 10^{6}$ K. The initial magnetic field configuration is a pseudostreamer arcade embedded in a uniform vertical background field (e.g., \citetalias{Edmondson2010b}; \citealt{Lynch2013}). The background field strength is $B_{0} = 3.0$~G. The characteristic plasma beta is $\beta_{0} \equiv 8 \pi P_{0} / B_{0}^{2} = 0.1$, and characteristic Alfv\'{e}n speed $V_{A0} \equiv B_{0} / \sqrt{4 \pi \rho_{0}} = 650$~km~s$^{-1}$. The characteristic length scale is $L_0 = 10^{9}$~cm, which yields a characteristic Alfv\'{e}n time of $\tau_{A0} \equiv L_{0} / V_{A0} = 15.385$~s. Furthermore, we define a characteristic current density magnitude, $J_{0} = (c/ 4 \pi) \ B_{0} / L_{0} = 7.162$ statamp~cm$^{-2}$.

We investigate the effect of increasing the guide field strength on the self-consistent current sheet formation, evolution, and magnetic reconnection dynamics. We define a guide field factor normalized to the uniform background field strength, $\zeta \equiv B_{GF}/B_{0}$, and use this notation to describe, respectively, the zero ($\zeta = 0.0$), weak ($\zeta = 0.1$), moderate ($\zeta = 0.5$), and strong ($\zeta = 1.0$) guide field cases. The $\zeta$ parameter spans the transition from strictly anti-parallel reconnection to component reconnection, i.e., $\zeta = \tan \left( \phi \right)$ for relative alignment angles of $\phi = \lbrace 0^{\circ}, 5.7^{\circ}, 26.6^{\circ}, 45^{\circ} \rbrace$. We note the variation in the guide field leads to variations in the respective system characteristics for each simulation: plasma beta $\beta_{\zeta} = \beta_{0} \left( \ 1 + \zeta^{2} \ \right)^{-1}$, Alfv\'{e}n speed $V_{A \zeta} = V_{A0} \left( \ 1 + \zeta^{2} \ \right)^{1/2}$, and Alfv\'{e}n time $\tau_{A \zeta} = \tau_{A0} \left( \ 1 + \zeta^{2} \ \right)^{-1/2}$.

\begin{figure*}
	\includegraphics[width=0.5\textwidth]{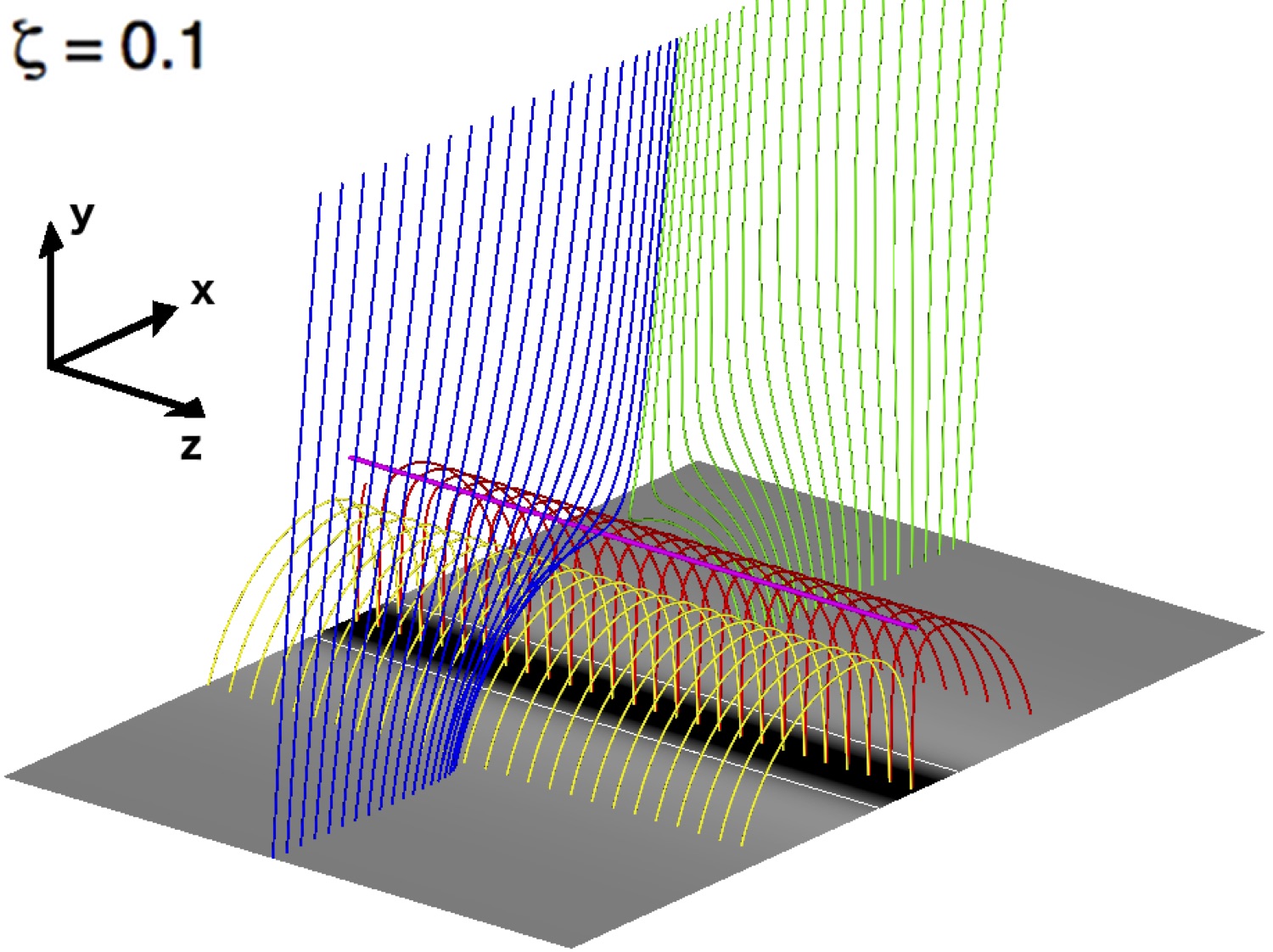}
	\includegraphics[width=0.5\textwidth]{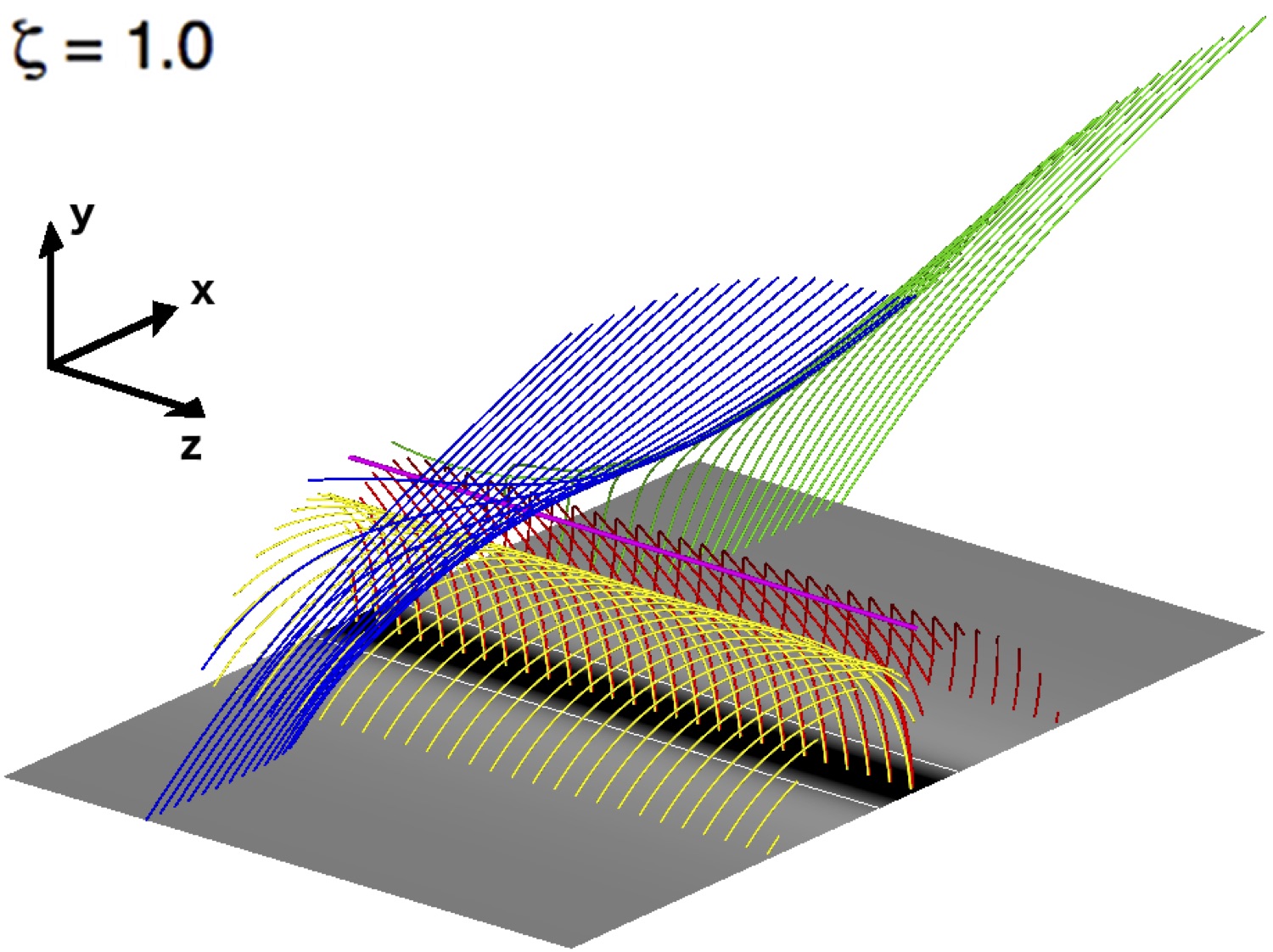}
	\caption{Representative field lines at $t = 0$ s showing the initial 3D magnetic 
	configuration of our translationally symmetric pseudostreamer arcade for weak 
	($\zeta = 0.1$, left panel) and strong ($\zeta = 1.0$, right panel) guide field 
	components. The $y=1$ surface shows the normal magnetic field component 
	$B_y(x,z)$ in grayscale at the lower boundary and the polarity inversion lines 
	are shown as the thin white lines under each arcade. The thick magenta field 
	line indicates the X-point null line.\\}
\label{F:InitGFImage}
\end{figure*}

The initial embedded pseudostreamer model is a translationally symmetric 
magnetic field, constructed from a the vector potential of the form
\begin{equation} \label{E:InitVecPot}
\bm{A}_{\zeta} = z B_{0} \left( \ \bm{\hat{x}} - \zeta \ \bm{\hat{y}} \ \right) + A_{z} (x,y) \ \bm{\hat{z}} \ .
\end{equation}
For all $\zeta$, the initial vector magnetic field 
$\bm{B}_{\zeta} = \nabla \times \bm{A}_{\zeta}$ is clearly independent of $z$. 
Furthermore, the underlying pseudostreamer model is constructed from the vector potential 
of a series of line dipoles at positions $\bm{r}_{n} = \pm nd \ \bm{\hat{x}}$,
\begin{equation} \label{E:VecPotZComp}
A_{z} ( x , y ) = M_{0} \ \sum_{n=0}^{N} \ \frac{ \left( \ \left( x \pm nd \right)/L_{0} \ \right) }{ \left( \ \left( x \pm nd \right)/L_{0} \ \right)^{2} + \left( \ y /L_{0} \ \right)^{2} } \ .
\end{equation}
The linear-dipole strength is set $M_{0} = -40$ G cm in order to produce a 
$B_{y}$ value of -25 G at the lower boundary of $y = 1.0 L_{0}$ in the 
pseudostreamer arcades. The dipole spacing length, $d = 15 L_{0}$, means 
the contributions of the line dipoles exactly cancel at $x=\pm n d / 2$ for an 
infinite series (and are effectively zero when the series is truncated at 
$N = 25$, see \citetalias{Edmondson2010b} for details).
Figure \ref{F:InitGFImage} shows an isometric perspective of the magnetic 
field structure for the $\zeta = 0.1$ (left panel) and $\zeta = 1.0$ (right panel) 
simulations at $t = 0$ which illustrates the tilt added in the $z$-direction by 
the guide field component. The closed pseudostreamer arcade flux systems 
are indicated by yellow and red field lines whereas the external unipolar open 
flux system is represented by blue and green field lines. The internal spine line 
(fan surface) is between the yellow and red flux systems whereas the external 
spine line (fan surface) is between the blue and green flux systems. 
Figure~\ref{F:GridAMR1}(a) shows the 2D $x$-$y$ projection of these flux 
systems which are identical in each simulation.

The system is energized in two stages with a spatially-uniform flow applied at 
the {\em upper boundary}, a distance of approximately 7$L_{0}$ away from the X-point. 
The first stage is a smooth acceleration phase 
followed by a constant flow for the duration of the simulations. The boundary 
flow is given by
$\bm{V}_{\rm{flow}} = 0.015 V_{A0} \ T(t) \ \hat{\bm{x}}$, 
where $T(t)$ is a piecewise-continuous, dimensionless function defined by
$T(t) = 1 - \cos{\left( \pi t / \left( 65 \ \tau_{A0} \right) \right)}$ for $0 \le t \le 65 \ \tau_{A0}$ 
and 
$T(t)=1$ for $t > 65 \ \tau_{A0}$. 
This corresponds to a uniform driving flow of $V_{0} = 20$~km~s$^{-1}$ for $t \ge 10^{3}$ s. 
We note that while the boundary flows are greater than typically observed photospheric 
velocities ($\sim$1~km~s$^{-1}$), our boundary flows remain extremely sub-Alfv\'{e}nic 
($M_{A} \sim 0.03$) and sub-sonic ($M_{s} \sim 0.10$). 

We employ periodic boundary conditions in the $x$- and $z$-boundaries of our Cartesian domain. In addition, we set extrapolative and zero-gradient boundary conditions at the upper and lower $y$-surfaces. Furthermore, we fix the lower $y$-boundary to no slip and reflecting. Despite the periodicity in the $x$- and $z$-boundary conditions, we achieve a quasi-steady state response such that the inflows at the reconnection layer reach an asymptotic value dictated by the constant boundary driving boundary flow.

\subsection{Computational Grid, Adaptive Mesh Refinement, \& Numerical Resistivity} \label{SS:Grid}

The computational domain is $x / L_{0} \in \left[ \ -7.5, 7.5 \ \right]$, 
$y / L_{0} \in \left[ \ 1.0 , 11.0 \ \right]$, and $z / L_{0} \in \left[ \ -5.0 , 5.0 \ \right]$. 
The computational grid is comprised of $3 \times 2 \times 2$ root blocks with $8^{3}$ 
grid cells per block. There are 6 levels of static grid refinement and two additional 
levels of dynamic refinement. The maximum level 8 grid resolution corresponds to 
an effective resolution of $3072 \times 2048 \times 2048$ and is equivalent to the 
spatial resolution used in \citetalias{Edmondson2010b}. The length of the 
computational grid cell in the highest-refinement region is thus 
$\delta = 10L_0/2048 = 4.883 \times 10^{6}$ cm.

\begin{figure*}
	\centerline{\includegraphics[width=1.0\textwidth]{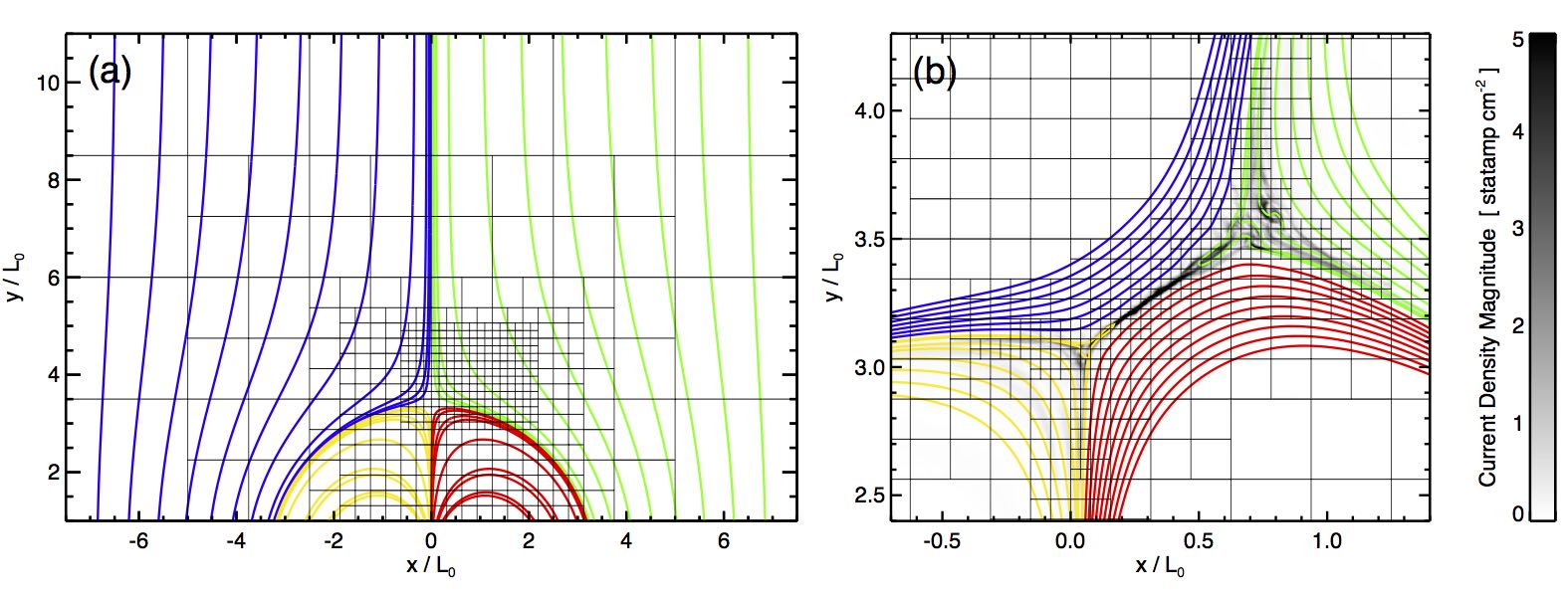}}
	\caption{(a) Block structure of the six levels of static grid refinement through 
	all $z =$ constant planes (with representative magnetic field lines at $t = 0$). 
	(b) Current density magnitude $|J|$ through the $z = 0$ plane at $t = 104\tau_{A0}$ 
	in simulation $\zeta = 0$ showing the two additional levels of dynamic block 
	refinement. \\
}
\label{F:GridAMR0} \label{F:GridAMR1}
\end{figure*}

Figure \ref{F:GridAMR0} panel (a) shows the root block structure of the static grid 
over the full domain at $t = 0$ with representative magnetic field lines colored as in 
Figure~\ref{F:InitGFImage}. 
The static grid structure is translationally symmetric through all $z =$ constant planes, 
for all $\zeta$ simulations. The adaptive mesh refinement (AMR) is conditional to the 
current density magnitude $|J|$. We fix the non-dimensional AMR refinement, $J_{l} / J_{0}$, 
and derefinement, $J_{r} / J_{0}$, thresholds to $7.0 \times 10^{-7}$ and $4.2$, respectively. 

%
Figure \ref{F:GridAMR1} panel (b) shows the current density magnitude $|J|$ in grayscale
through the $z = 0$ plane at $t = 104 \ t_{A0}$ for the no guide-field $\zeta = 0$ case. The 
dynamic grid refinement region is concentrated at the current sheet where the blue and red 
magnetic flux is reconnecting to become part of the yellow and green flux systems. 

We solve the equations of ideal MHD without an explicit, physical magnetic resistivity. However, necessary and stabilizing numerical diffusion terms introduce an effective viscosity and resistivity on very small spatial scales. In this way, magnetic reconnection occurs when the large magnetic field gradients associated with current sheet features become under-resolved, i.e., have been compressed to the local grid resolution scale. One usually estimates the effective numerical resistivity by balancing the average inflow timescale of magnetic flux into the current sheet (of thickness $\delta$) against the resistive dissipation 
timescale,
\begin{equation} \label{E:NumResistivity}
\eta \sim \delta \ \langle \ V_{in} \ \rangle \ .
\end{equation}
We note that Equation~(\ref{E:NumResistivity}) is a simple balance between the ideal transverse and dissipation timescales, and therefore should be considered an over-estimate of the actual numerical resistivity. For ARMS, and finite-volume FCT schemes in general, the numerical resistivity is a \emph{local} quantity that depends explicitly on the gradients of the field and flows used to determine how much of the higher-order anti-diffusion correction is applied to the lower-order diffusive transport \citep{DeVore1991}. While Equation~(\ref{E:NumResistivity}) is the estimate typically used in MHD simulations there remains a question as to why many numerical simulations with Lundquist numbers of a few~$\times 10^3$ (e.g. \citetalias{Edmondson2010b}; \citealt{Ni2010}; \citealt{Shen2011}; \citealt{Lynch2016a}), exhibit nonlinear dynamics and plasmoid-unstable evolution similar to those simulations with higher Lundquist numbers that exceed the more traditional $\gtrsim 10^4$ threshold \citep[e.g.][]{Furth1963,Biskamp1986,Huang2010,Loureiro2012,Murphy2013}. 

In Appendix \ref{SB:NumRes}, we derive corrections to the Equation~(\ref{E:NumResistivity}) numerical resistivity estimate based on the transport of total energy density through the current sheet to obtain
\begin{equation} \label{E:NumResistivityCorretion}
\eta \sim \delta \ \langle \ V_{in} \ \rangle \ \frac{\left( \ \chi + Z \ \right)}{4} ,
\end{equation}
where the dimensionless factors $\chi$ and $Z$ describe the advection of magnetic energy 
density and plasma compressibility, respectively. To first order in the current sheet aspect ratio $\delta / \langle \ L_{CS} \ \rangle$,
\begin{equation} \label{E:AdvectionFactor}
\chi \equiv 1 - \left( \ 2 \ \frac{\langle \ B_{out} \ \rangle}{\langle \ B_{in} \ \rangle} + \frac{\langle \ V_{out} \ \rangle}{\langle \ V_{in} \ \rangle} \ \frac{\mathcal{ME}_{out}}{\mathcal{ME}_{in}} \ \right) \frac{\delta}{\langle \ L_{CS} \ \rangle} \ ,
\end{equation}
\begin{equation} \label{E:CompressibilityFactor}
Z \equiv M_{Ain}^{2} \left[ \ 1 - \left( \ 2 \ \frac{\langle \ B_{out} \ \rangle}{\langle \ B_{in} \ \rangle} + \frac{\langle \ V_{out} \ \rangle}{\langle \ V_{in} \ \rangle} \ \frac{\mathcal{KE}_{out}}{\mathcal{KE}_{in}} \ \right) \frac{\delta}{\langle \ L_{CS} \ \rangle} \ \right] \ .
\end{equation}
The brackets signify an appropriate spatial-averaging procedure, such that $\mathcal{ME} = \langle \ B \ \rangle^{2} / (8 \pi)$ and $\mathcal{KE} = \langle \ \rho \ \rangle \langle \ V \ \rangle^{2} / 2$ are the respective magnetic and kinetic energy densities advected with the inflows and outflows, and $M_{Ain} = \langle \ V_{in} \ \rangle / \langle \ V_{Ain} \ \rangle$ is the Alfv\'{e}nic Mach number of the inflow. From Equation (\ref{E:NumResistivityCorretion}), one sees immediately that the traditional $\eta \sim \delta \ \langle \ V_{in} \ \rangle$ numerical resistivity estimate could be overestimated by up to a factor of 4 in the thin current sheet, incompressible (low Alfv\'{e}nic Mach number) inflow limit ($\chi \rightarrow 1$, $Z \rightarrow 0$).

\section{Simulation Results} \label{S:Results}

The basic evolution of our system is that of driven reconnection in which the reconnection inflow speed adjusts in an attempt to achieve a quasi-steady state where the rate of material and flux processed through the current sheet is determined by the distant boundary driving flows. For very slow driving, the reconnection is essentially the Sweet-Parker type and the current sheet remains laminar in the sense of a quasi-static, reversible thermodynamic process. As the driving increases, the system attempts to process more material and flux through the current sheet and the reconnection rate necessarily increases. Once the rate of material and magnetic flux to process exceeds the current sheet's capacity via the Sweet-Parker type reconnection, basic mass and energy continuity necessitate the rapid onset of nonlinearity and the development of plasmoid-unstable evolution \citep[see also][]{Daughtonetal2011a,Daughtonetal2011b,PucciVelli2014,Teneranietal2016}.

\subsection{Onset of the Non-Linear Phase of the Plasmoid Instability} \label{SS:NonLinOnset}

The driving flows at the upper boundary increases the magnetic energy such that the material and 
flux are being processed/transferred through the current sheet at the same rate. Magnetic flux is 
transferred so as to spread the volumetric (force-free) currents into as large a volume as possible. For relatively weak 
guide field cases, the magnetic energy is (marginally) reduced as flux is processed through the 
current sheet and transferred between domains; that is until the overlying field becomes stretched 
enough to saturate this effect throughout the entire box. Increasing the guide field has the effect of 
suppressing this process/transfer effect through the current sheet. In other words, a stronger guide 
field smooths the magnetic field gradients which delays the onset of nonlinear dynamics. We note 
this system is not undergoing free-reconnection, but rather, the reconnection dynamics are driven 
in a well-controlled quasi-steady state manner such that the inflow Alfv\'{e}nic Mach number matches that of the driving 
flows at the boundary. 

\begin{figure*}
	\includegraphics[width=3.6in]{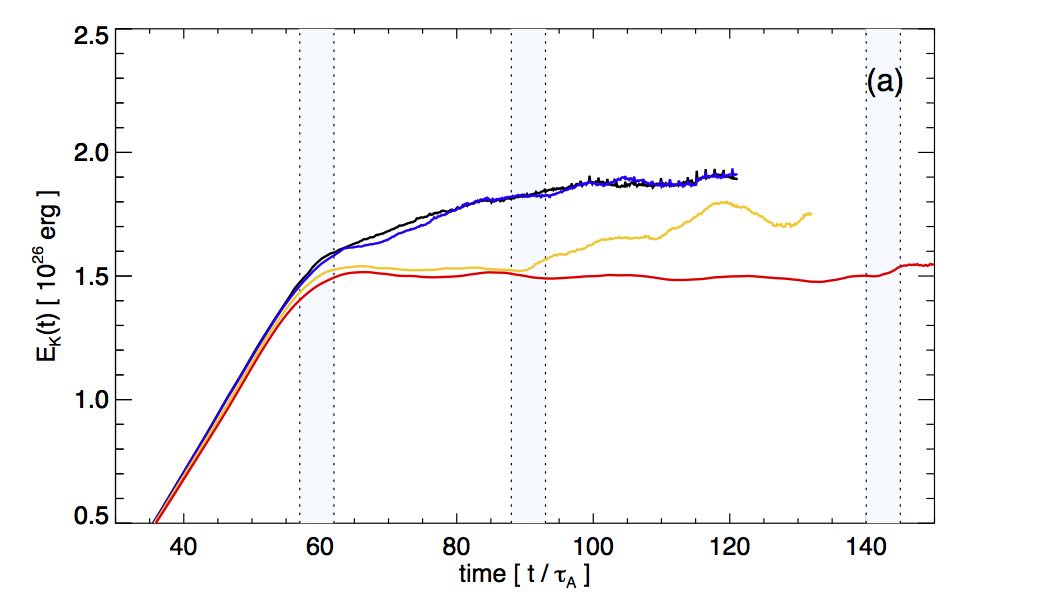}  
	\includegraphics[width=3.6in]{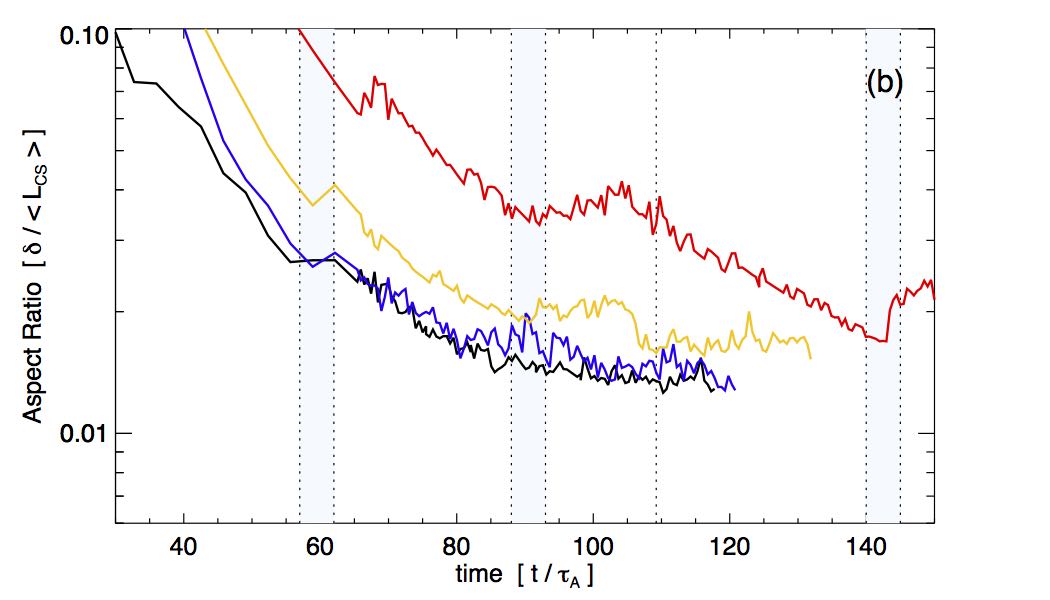}
	\includegraphics[width=3.6in]{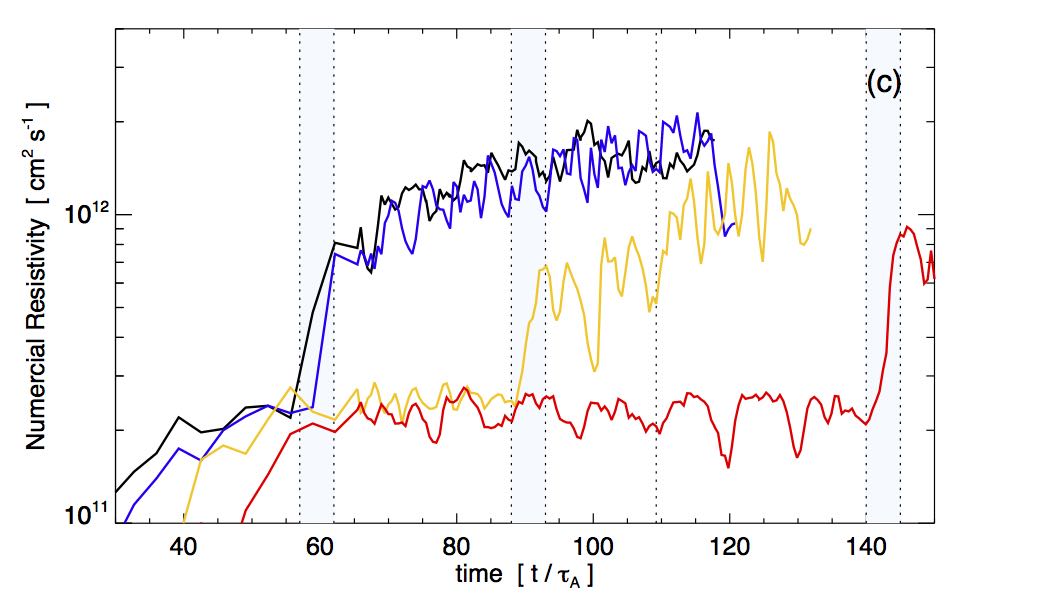}    
	\includegraphics[width=3.6in]{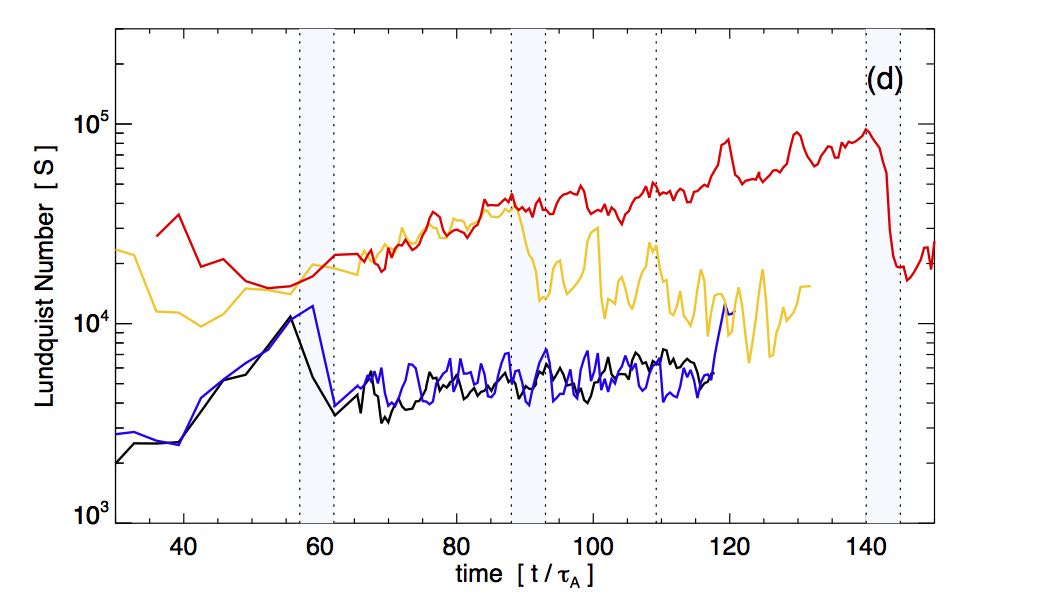}
	\caption{(a) Kinetic energy evolution. (b) Current sheet aspect ratio evolution. 
	(c) Numerical resistivity evolution. (d) Lundquist number evolution. 
	Every panel shows the four different guide field simulations: black is 
	$\zeta = 0.0$, blue is $\zeta = 0.1$, yellow is $\zeta = 0.5$, and red is 
	$\zeta = 1.0$. Vertical blue stripes indicate the onset of nonlinear evolution 
	and the plasmoid instability.}
\label{CSprop}
\end{figure*}

The delay in the onset of the nonlinear reconnection phase can be seen in Figure~\ref{CSprop}(a) which shows the global kinetic energy in the simulation domains for all four simulations: $\zeta = 0.0$ (black), $\zeta = 0.1$ (blue), $\zeta = 0.5$ (yellow), and $\zeta = 1.0$ (red). The total kinetic energy of the uniform driving at the upper boundary can be seen to be approximately $1.5\times10^{26}$~erg. Any additional kinetic energy is associated with the plasmoid-unstable reconnection dynamics and relatively fast reconnection jet outflow. The zero and weak, $\zeta = 0.0$ and $0.1$, guide field cases (black, blue lines) immediately transition into the nonlinear phase by $t / \tau_{A} \gtrsim 56$. The moderate, $\zeta = 0.5$, guide field simulation (yellow line) begins this transition by $t / \tau_{A} \gtrsim 88$ and the strong, $\zeta = 1.0$, guide field case (red line) only begins to show signs of this transition at $t / \tau_{A} \gtrsim 140$. These transitions are marked as the vertical, light blue stripes in every panel of Figure~\ref{CSprop}.

Figure~\ref{CSprop}(b) shows that the current sheet aspect ratio, ${\delta}/{\langle \ L_{CS} \ \rangle}$, decreasing smoothly through the nonlinear onset timescales, beyond which it settles to order $10^{-2}$. In all cases, the smooth current sheet formation prior to the non-linear onset timescale corresponds to the opening of a Syrovatskii-type current sheet. We note, the linear-phase fluctuations in all panels of Figure \ref{CSprop} are an effect of the current sheet length identification even though the current sheet itself is relatively laminar; beyond the non-linear onset, the fluctuations grow in amplitude due to plasmoid formation, ejection, and current sheet re-formation.

\begin{figure*}
	\includegraphics[width=3.6in]{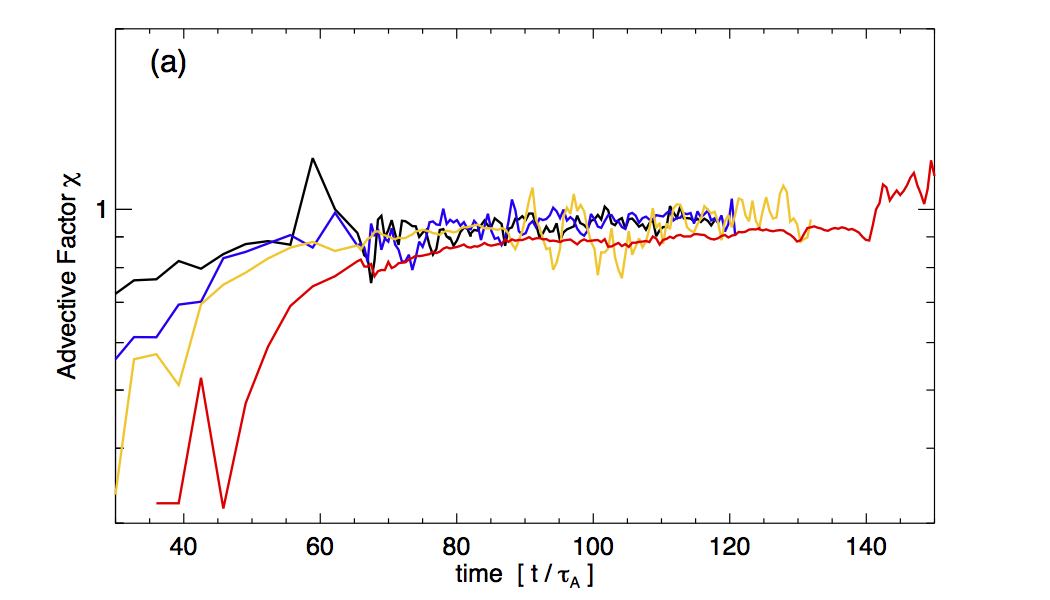}  
	\includegraphics[width=3.6in]{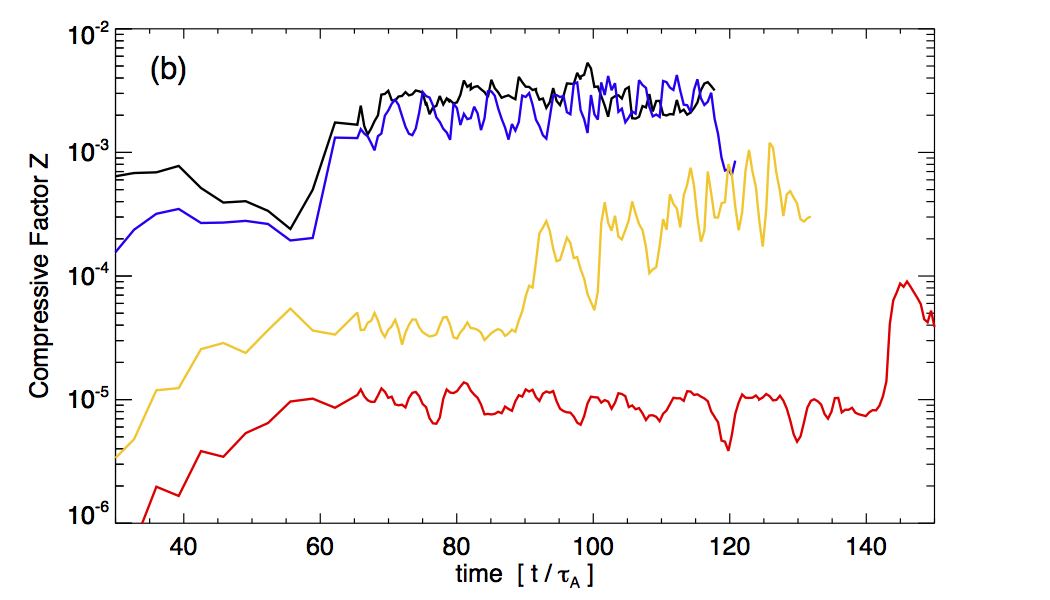}
	\caption{(a) Advective correction factor $\chi$. (b) Compressive correction factor $Z$.}
\label{NumResCorrections}
\end{figure*}

Figure~\ref{CSprop}(c) shows the evolution of our improved numerical resistivity estimate, Equation~(\ref{E:NumResistivityCorretion}). In all cases, the advection and compressibility correction terms are, respectively, $\chi \sim 1$ and $Z \lesssim 10^{-2}$ (see Figure \ref{NumResCorrections}). During the initial spine line separation and current sheet formation stage (once a resistivity estimate becomes well-defined) we see a consistent numerical resistivity of approximately $\eta \sim 2 \times 10^{11}$~cm$^{2}$~s$^{-1}$ independent of the guide field strength. The onset of nonlinear dynamics coincide with a relatively rapid increase in the estimated numerical resistivity; a phenomenology consistent with \citet{Cassaketal2007a}. Again, this nonlinear onset discontinuity in the resistivity estimate mirrors the global kinetic energy evolution: around approximately $t / \tau_{A} \sim 57 - 62$ for the weak guide field ($\zeta = 0.0, 0.1$) cases, $t / \tau_{A} \sim 88 - 93$ for the moderate guide field ($\zeta = 0.5$) case, and between $t / \tau_{A} \sim 140 - 145$ for the relatively strong guide field ($\zeta = 1.0$) case. After nonlinear evolution onset, the numerical resistivity estimates converge to order $\eta \sim 10^{12}$~cm$^{2}$~s$^{-1}$.

In Figure~\ref{CSprop}(d) we show estimate the local Lundquist number in the 
vicinity of the current sheet given by,
\begin{equation} \label{LocalLundquist}
S \simeq \frac{\langle \ L_{CS} \ \rangle \ \langle \ V_{Ain} \ \rangle}{\langle \ \eta \ \rangle} = \frac{4 \left( \ 1 + \zeta^{2} \ \right)^{1/2}}{M_{Ain}} \ \frac{\langle \ L_{CS} \ \rangle}{\delta}.
\end{equation}

\noindent In all cases, since the dissipation scale $\delta$ is fixed by the gird resolution, 
the Lundquist estimate increases as the current sheet opens, i.e., linearly with both 
$\langle \ L_{CS} \ \rangle$ growth, as well as with increasing guide field strength, 
$\vert \zeta \vert$. The linear dependence on the current sheet length is consistent 
with Figure~\ref{CSprop}(b) (on the semi-log scale) prior to the jump discontinuity at 
nonlinear onset. Furthermore, the guide field strength dependence is evident in the 
approximate ordering of the curves. With the improved numerical resistivity estimate 
of Equation (\ref{E:NumResistivityCorretion}), the local Lundquist number beyond 
nonlinear onset converges to approximately $S \sim 10^{4}$.

\begin{figure*}
	\includegraphics[width=1.0\linewidth,]{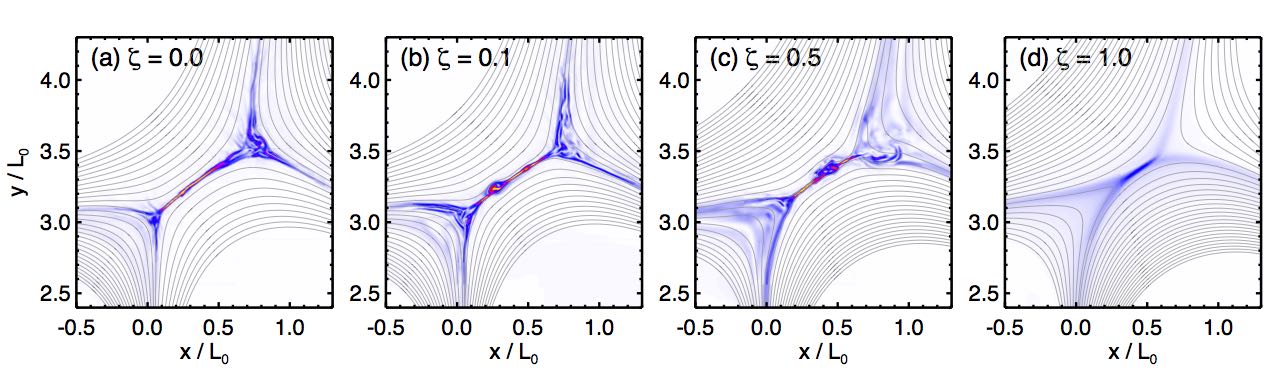}
	\caption{The current sheet development in the $z=0$ transverse planes in each of 
	the simulations at $t = 117.79 \ \tau_A$ shows the delay of the onset of non-linear 
	dynamics of the plasmoid instability with increasing guide field strength. Panel (a) 
	$\zeta = 0.0$; panel (b) $\zeta = 0.1$; panel (c) $\zeta = 0.5$; and panel (d) $\zeta = 1.0$.}
\label{CSTransversePlane}
\end{figure*}

In all simulation cases, the nonlinear onset time is ordered by the magnitude of the guide field, as expected, since the inclusion of the guide field reduces the overall magnitude of the magnetic field gradients in the vicinity of the diffusion region. This effect is also evident in Figure~\ref{CSTransversePlane} which shows the $z = 0$ transverse plane of $|J|$ with representative field lines for each current sheet simulation at $t/\tau_{A} = 117.79$. It is clear that the strong guide field case, panel (d) of Figure~\ref{CSTransversePlane}, has not yet entered the nonlinear evolution, plasmoid-generation regime.

\subsection{Qualitative Guide Field Influence on Reconnection-Generated Structure} \label{SS:FluxRopeForm}

\begin{figure*}[!t]
	\centerline{\includegraphics[width=0.95\textwidth]{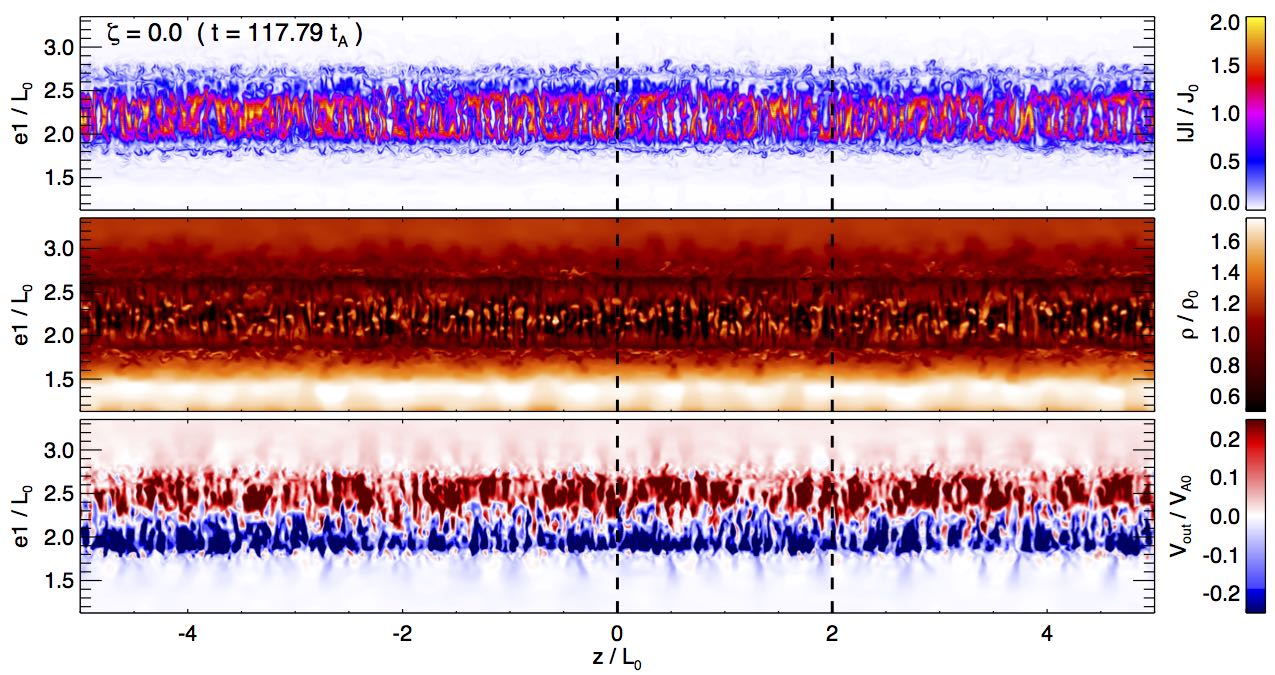}}
	\caption{Zero guide field case ($\zeta = 0.0$) at $t = 117.79 \ \tau_A$. 
	Upper panel: current density magnitude $|J|/J_{0}$ (top), mass density 
	ratio $\rho/\rho_0$ (middle), and outflow velocity $V_{out}/V_A$ 
	(bottom). \\
	(An animation of this figure is available) 
	}
\label{ZeroGF}
\end{figure*}

\begin{figure*}
	\centerline{\includegraphics[width=0.95\textwidth]{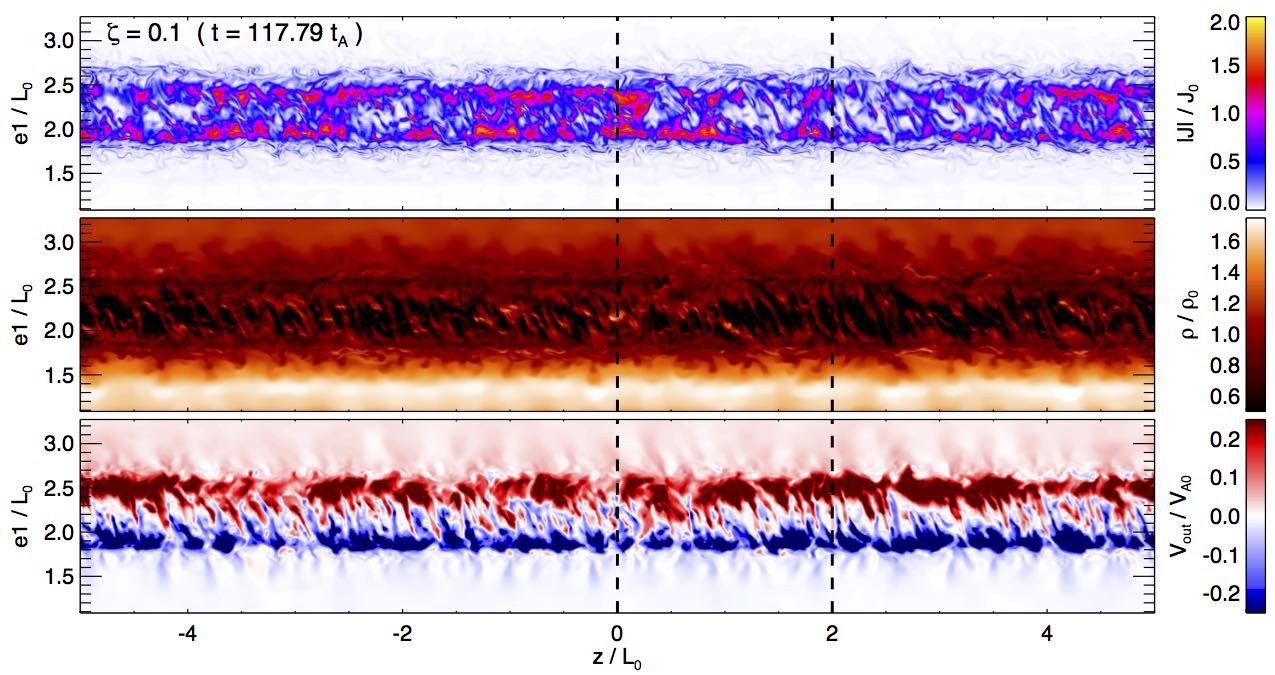}}
	\caption{Weak guide field case ($\zeta = 0.1$) at $t = 117.79 \ \tau_A$ in 
	the same format as Figure~\ref{ZeroGF}. \\
	(An animation of this figure is available) 
	}
\label{WeakGF}
\end{figure*}

\begin{figure*}
	\centerline{\includegraphics[width=0.95\textwidth]{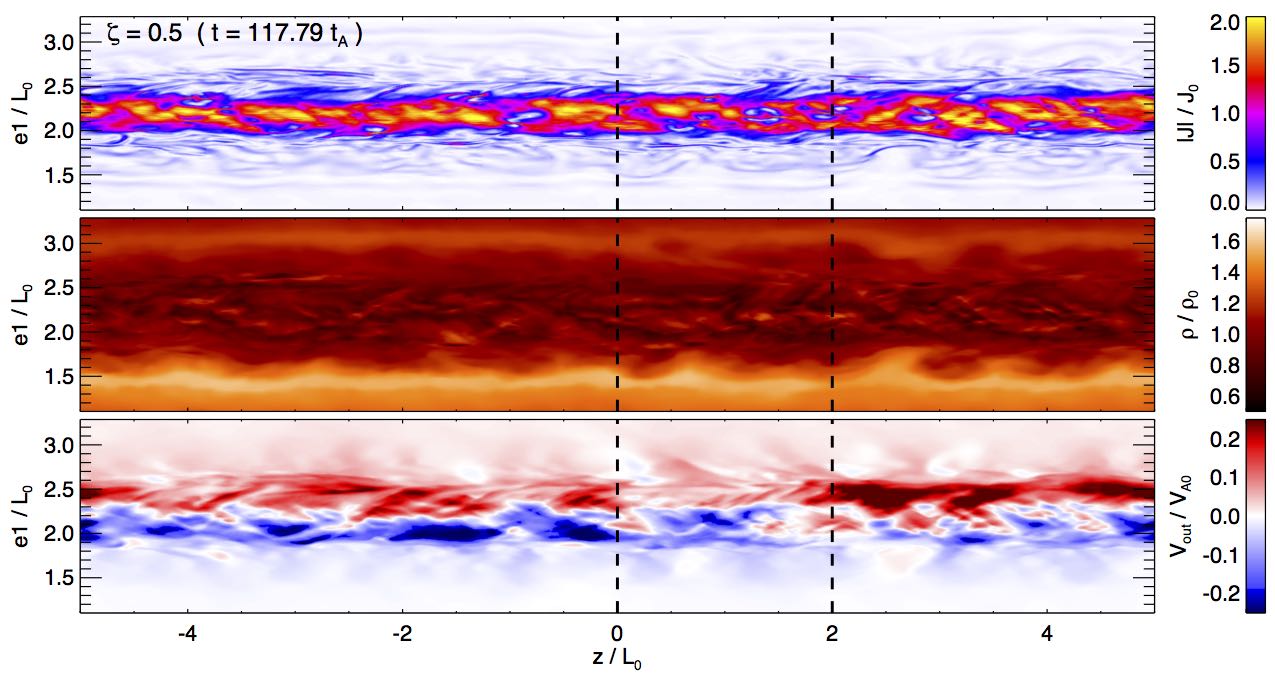}}
	\caption{Moderate guid field case ($\zeta = 0.5$) at $t = 117.79 \ \tau_A$ in 
	the same format as Figure~\ref{ZeroGF}. \\
	(An animation of this figure is available) 
	}
\label{MdrtGF}
\end{figure*}

\begin{figure}
	\centerline{\includegraphics[width=0.9\textwidth]{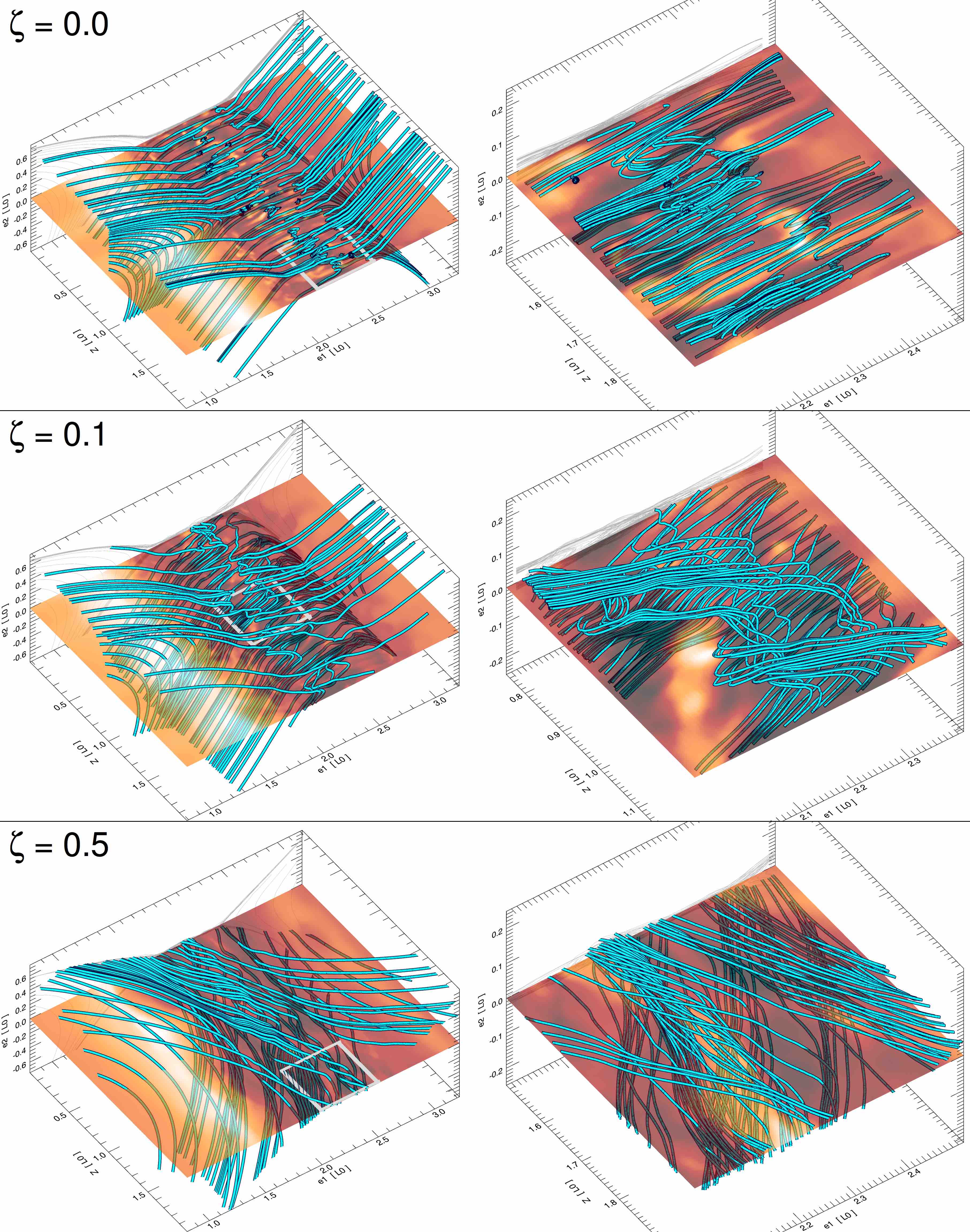}}
	\caption{Three-dimensional view of representative 
	magnetic field lines through mass density ratio $e_2=0$ plane for 
	$z/L_0 \in \left[ 0, 2 \right]$ from Figures~\ref{ZeroGF}--\ref{MdrtGF}. 
	Top panel: $\zeta = 0.0$; Middle panel: $\zeta = 0.1$; Bottom panel: $\zeta = 0.5$. \\ } 
\label{FLICx}
\end{figure}

Qualitatively, the general picture is that the guide field not only delays the dynamics by reducing the field gradients in the vicinity of the current sheet, but also increases the structure coherency in the same direction. We illustrate this in Figures~\ref{ZeroGF}--\ref{FLICx} at $t = 117.79 \ \tau_A$ (consistent with Figure~\ref{CSTransversePlane}) for the zero, weak, and moderate guide field cases ($\zeta = 0$, $\zeta = 0.1$, and $\zeta = 0.5$) which we describe below. We note, the strong $\zeta = 1.0$ guide field case has not yet transitioned to non-linear evolution. Figures~\ref{ZeroGF}--\ref{MdrtGF} each have their corresponding animations included as electronic supplements to the online version of the article. 

In Figure~\ref{ZeroGF}, upper panel, we plot properties of the $\zeta=0$ current sheet in local, current-sheet centered coordinates: top row, intensity of the current density; middle row, the mass density ratio relative to the initial state; and bottom row, the normalized outflow velocities. The guide field direction is along $\hat{z}$ and the $\hat{e}_{1}$ direction is parallel to the current sheet (see Appendix~\ref{SA:CSFit} for details). The perspective is observing the current sheet dynamics from the top left corner of the panels of Figure~\ref{CSTransversePlane}. The main result from \citetalias{Edmondson2010b} was that, for the $\zeta = 0$ case, the magnetic reconnection occurs in different transverse planes and proceeds essentially independently from one another. Specifically, there was almost no longitudinal ($\hat{z}$-direction) coherency in the reconnection-generated density enhancements and field structure within the current sheet. This is seen in Figures~\ref{ZeroGF} as the collimation of the current density, mass density, and outflow velocities along the $\hat{e}_{1}$-direction. 

Figure~\ref{WeakGF} shows the $\zeta = 0.1$ guide field case 
in the same format as Figures~\ref{ZeroGF}.
This case has more coherent structure in the $z$-direction, despite being almost 
identical to the $\zeta = 0$ case in the global reconnection properties (c.f. properties 
in Figure~\ref{CSprop}). We note that \citet{Lynch2016a} discussed how even 
a relatively weak guide field in the reconnection inflow regions can generate a broader 
distribution of localized, stronger concentrations of guide field (and current and mass 
densities) during plasmoid formation in plasmoid-unstable current sheets. 

Figure~\ref{MdrtGF} shows the $\zeta = 0.5$ guide field case in the same format as 
Figures~\ref{ZeroGF} and \ref{WeakGF}. The moderate guide field simulation generates 
significantly more extended current density, mass density, and reconnection jet outflow 
structure along the $\hat{z}$-direction of the current sheet. 

Figure~\ref{FLICx} shows 3D views of representative magnetic 
field lines in the $z / L_0 \in [0, 2]$ subvolume centered on the current sheet. 
The semi-transparent $e_2=0$ planes in each panel shows the mass density ratio 
within the vertical dashed lines of Figures~\ref{ZeroGF}--\ref{MdrtGF}. 
Figure~\ref{FLICx} visually confirms that the current sheet mass density enhancements 
are signatures of 3D magnetic islands \citep[qualitatively similar to][Figures 1a and 4]{Daughtonetal2011a}
Figure~\ref{FLICx}, top panel, shows that the $\zeta = 0$ twisted field lines characteristic 
of 3D magnetic island structures in the guide field case have an approximate 
longitudinal coherency of $\delta / L_0 \sim 10^{-2}$, corresponding to approximately 
2--3 grid planes---roughly the width of the plasmoid density enhancement in the 
transverse plane. Furthermore, the largest $\zeta = 0$ structure coalescence 
grows to approximately $0.1 L_{0}$ before ejection. 
Figure~\ref{FLICx}, middle panel, shows multiple, coherent twisted flux ropes on the 
order of $\sim 0.5 L_{0}$ in length in the current sheet in the $\zeta = 0.1$ case. 
Figure~\ref{FLICx}, bottom panel, shows that the 3D magnetic island flux 
rope coherency in the $\zeta = 0.5$ case extends even further, reaching lengths on the 
order of $\sim 1 - 2 L_{0}$ in the guide field direction.

\subsection{Quantifying Reconnection-Generated Coherency Scales Due to the Guide Field}
\label{SS:CharCoherScl}

We quantify the coherency scales associated with the plasmoid density fluctuations 
generated during the nonlinear phase of magnetic reconnection and use these scales to analyze 
the Fourier power spectra. While our system is not ``turbulent'' per se, we will show that 
the statistical and spectral properties of the reconnection-generated density fluctuations in 
our current sheets can be understood in terms of the traditional turbulence regime 
classifications of energy injection and continuation range, inertial range, and dissipation 
scale. The coherency scales of our plasmoid formation and evolution are defined by the 
following transitions between these regimes:
\begin{enumerate}
\item the \emph{minimum structure formation coherency scale} is the transition between the 
energy injection (structure formation) range, and the so-called energy continuation range; 
\item the \emph{maximum inertial range scale} is the transition between the energy 
continuation range and the traditional turbulent inertial range; and 
\item the \emph{dissipation scale} is set by the maximum numerical resolution of 
our MHD simulations.
\end{enumerate}

\noindent In each transition, there is a marked break between spectral indices.

During the non-linear evolution phase, the AMR covers the current sheet with the highest 
level of grid refinement. Since there is no physical dissipation scale imposed on our simulations, 
we take the numerical dissipation scale as three times the maximum grid resolution, 
$\xi_{D} = z_{D} / L_{0} = 3 \ \delta / L_{0} = 1.5 \times 10^{-3}$.

\subsubsection{Minimum Structure Formation Coherency Scale} \label{MinStrFormScl}

Wavelet methods are commonly used to characterize structure coherency scales in 
non-steady signals \citep[e.g., see][and references therein]{Edmondson2013b}. In this 
analysis, we employ the Morlet wavelet transform of the normalized mass density, 
$\rho(z)/\rho_{0}$, given by,
\begin{equation} \label{E:WaveletTransform}
W_{\rho}( z , \xi ) = \int \frac{\rho ( z' )}{\rho_{0}} \ \psi^* ( z' , z , \xi ) \ dz' \ ,
\end{equation}
\begin{equation}
\psi \left( z' , z , \xi \right) = \frac{\pi^{1/4}}{\vert \xi \vert^{1/2}} {\rm exp}\left[ i \omega_{0} \left( \frac{z' - z}{\xi} \right) \right] {\rm exp}\left[ - \frac{1}{2} \left( \frac{z' - z}{\xi} \right)^2 \right] \ ,
\end{equation}

\noindent where $z'$, $z$ are spatial coordinates in the longitudinal direction along 
the current sheet, $\xi$ is therefore the longitudinal spatial coherency scale, and 
$\omega_{0}$ is a fixed non-dimensional frequency parameter (set to $\omega_{0} = 6$). 
The integrated power per scale (IPPS), obtained from the rectified wavelet power spectrum as
\begin{equation}
{\rm IPPS}(\xi) = \int dz \ \vert \xi \vert^{-1} \ \vert W_{\rho} \left( z , \xi \right) \vert^2,
\end{equation}

\noindent is a modified proxy measure of the Fourier power. The rectification is 
necessary due to a bias in favor of large timescale features in the canonical power 
spectrum, which are attributed to the width of the wavelet filter in frequency space 
(at large timescales the function is highly compressed yielding sharper peaks of 
higher amplitude, whereas at small timescales the wavelet filter is broad, 
underestimating the high frequency peaks). The modification from the standard 
Fourier power spectrum is due to the wavelet cone-of-influence, which acts as 
a high-pass filter, attenuating the lowest frequency structures.

\begin{figure}
\centerline{ \includegraphics[width=0.5\textwidth]{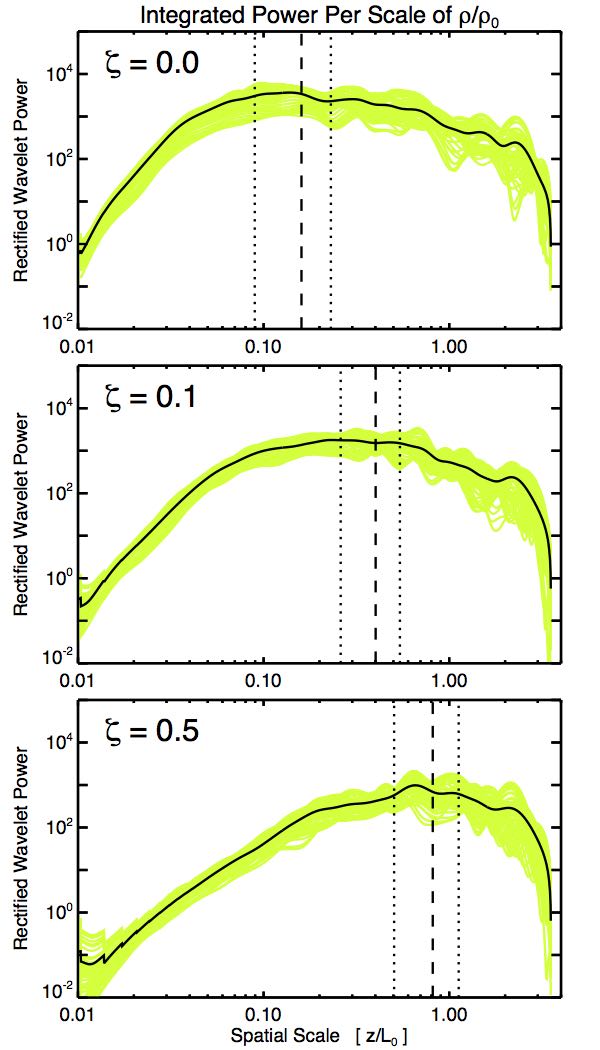}
		  \includegraphics[width=0.5\textwidth]{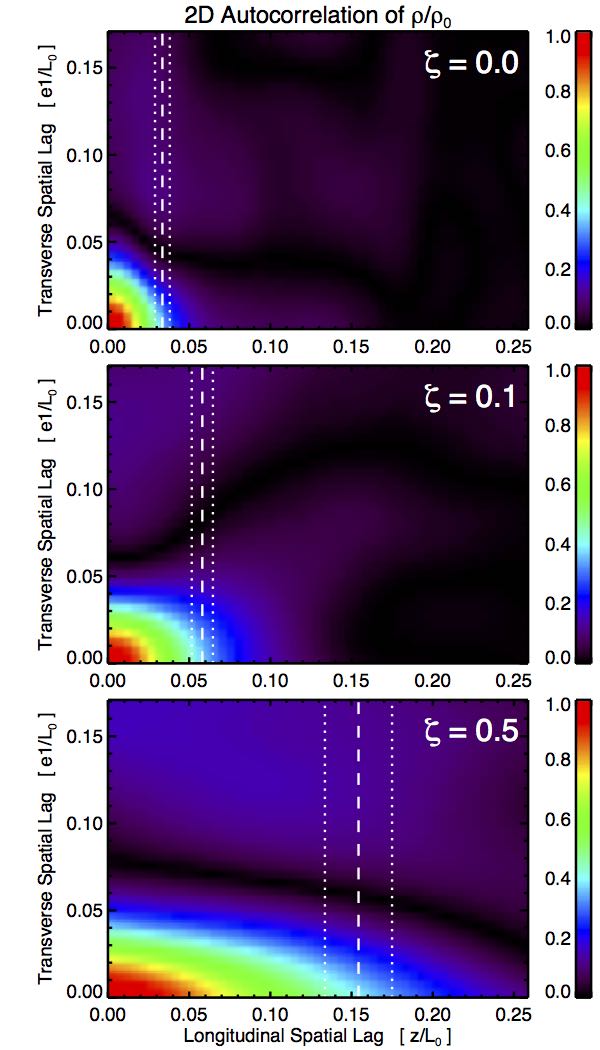} }
	\caption{Left column: Rectified wavelet integrated power per scale of 
	the density fluctuations in the current sheet for $\zeta = 0.0$ (top), $\zeta = 0.1$ (middle), 
	and $\zeta = 0.5$ (bottom) simulations. The minimum structure formation 
	coherency scales are indicated with the vertical dashed lines. Right column: 
	2D autocorrelation functions of plasmoid density fluctuations in each of the 
	CSs. The maximum inertial range scales are indicated with the vertical 
	dashed lines. \\}
\label{FigIPPS2DAC}
\end{figure}

At each time step, in all simulation cases, we calculate the IPPS$(\xi)$ for every one-dimensional longitudinal mass density slice, $\rho(z)/\rho_{0}$, within the current sheet at that time step. The left column of Figure~\ref{FigIPPS2DAC} plots the rectified wavelet IPPS$(\xi)$ spectra for each 1D-data array (green) within the current sheet at an sample time $t = 109.229 \ \tau_A$ for each $\zeta=\{ 0.0, 0.1, 0.5 \}$ simulation. From all the IPPS spectra in each current sheet at each time, we calculate a distribution of global maximum IPPS values. {\em We interpret the spatial coherency scale corresponding to the average global maximum of the set of IPPS spectra to be the minimum structure formation coherency scale $\xi_{I}$ with $\pm$1$\sigma_{I}$ uncertainty.} In canonical turbulence, the structure formation/energy injection range is all coherency scales $\gtrsim \xi_{I}$.

The key point shown in the left column of Figure~\ref{FigIPPS2DAC}, is that with 
increasing guide field strength the minimum structure coherency formation scale 
increases from $\sim$0.15 to $\sim$0.70. The minimum coherency formation 
scales $\xi_{I}$ and corresponding $\pm 1 \sigma_{I}$ uncertainty are tabulated 
for each guide field case $\zeta$ in Table \ref{Tbl1}. The shape of the spectra at 
smaller spatial scales (higher spatial frequencies) also shows the increasing 
development of a power-law character. 

\subsubsection{Maximum Inertial Range Scale} \label{MaxInertScl}

Statistical autocorrelation methods are commonly used to characterize coherency 
scales \citep{Zurbuchen2000}. In this analysis, we calculate the discrete two-dimensional 
autocorrelation function of the normalized mass density $\rho(\bm{r})/\rho_{0}$ sampled 
in the current sheet plane where $\bm{r}$ is the position vector in the $e_1$-$z$ plane. 
The discrete 2D autocorrelation function is given by
\begin{equation} \label{e:acf}
A(\lambda,\mu) = \frac{ \displaystyle \sum_{l=0}^{L-\lambda-1} \sum_{m=0}^{M-\mu-1}  \left( \ \rho_{l,m} / \rho_{0} - < \rho / \rho_{0} > \ \right) \left( \ \rho_{l+\lambda,m+\mu} / \rho_{0} - < \rho / \rho_{0} > \ \right) }{ \displaystyle \sum_{l=0}^{L-1} \sum_{m=0}^{M-1} \left( \ \rho_{l,m} / \rho_{0} - < \rho / \rho_{0} > \ \right)^2 } \ ,
\end{equation}
where $< \rho / \rho_{0} >$ is the mean density enhancement within the current sheet. 
The subscripts $l, m$ denote the value of the normalized density, $\rho_{l,m} / \rho_{0}$,
at the grid point with label $l$ in the $z$-direction and $m$ in the $e_{1}$-direction. 
$\lambda$ and $\mu$ are fixed discrete spatial lag values in the $z$- and $e_1$-directions, 
respectively; in other words, $\lambda = N \Delta z$ and $\mu = M \Delta e_{1}$ are 
integer ($N$ and $M$) multiples of the grid cell lengths $\Delta e_1 = \Delta z = \delta$.

At each time step, in all simulation cases, we calculate the 2D autocorrelation for the 
density enhancement $\rho(\bm{r}) / \rho_{0}$ within the current sheet at that time step. 
The right column of Figure~\ref{FigIPPS2DAC} illustrates the 2D autocorrelation 
contours of our reconnection-generated density fluctuations in the current sheets 
for the $\zeta = \left\{ 0.0, 0.1, 0.5 \right\}$ simulations at $t = 109.229 \tau_A$. 
{\em We interpret the the $1/e$-folding contour of the 2D autocorrelation to be the maximum 
inertial range scale, $\xi_{A}$, of the reconnection-generated plasmoid density fluctuations 
within the current sheet.} The longitudinal spatial lag $\xi_{A}$ is shown as the vertical dashed line 
with $\pm 1\sigma_{A}$ uncertainty range as vertical dotted lines. The maximum inertial scales $\xi_{A}$ and 
corresponding uncertainty are tabulated for each guide field case 
$\zeta$ in Table \ref{Tbl1}.

The key point shown in the right column of Figure~\ref{FigIPPS2DAC}, is that with 
increasing guide field strength the width of the longitudinal ($z$-direction) autocorrelation 
function increases significantly while the transverse ($e_1$-direction) autocorrelation 
width remains approximately constant ($\sim$~5 - 10~$\delta/L_0$). This result suggests 
that structure coherency reflects formation \emph{only} in the guide field direction, as opposed to 
oblique formation. This effect lends support to the idea that identified oblique structures are the 
consequence of more complex, 3D dynamics such as the secondary instability \citep[e.g.][]{Dahlburgetal2005}.

\subsubsection{Density Fluctuation Fourier Power Spectra} \label{SSS:FP}

Having defined the relevant spatial scales in the previous sections, we are now able 
to interpret the Fourier power spectrum in the canonical turbulence regimes of energy 
injection $10 L_0 \ge \xi \ge \xi_{I}$, energy continuation $\xi_{I} \ge \xi \ge \xi_{A}$, 
inertial range $\xi_{A} \ge \xi \ge \xi_{D}$, and dissipation. The $\xi_{I}$ scale is the 
\emph{minimum structure formation coherency scale} identified via the wavelet IPPS 
spectra ($\S$ \ref{MinStrFormScl}), the $\xi_{A}$ scale is the \emph{maximum inertial 
range scale} identified via 2D autocorrelation ($\S$ \ref{MaxInertScl}), and the $\xi_{D}$ 
is the \emph{dissipation scale} fixed at slightly larger than the grid resolution $3 \ \delta$.

\begin{figure*}
	\centerline{\includegraphics[width=1.0\textwidth]{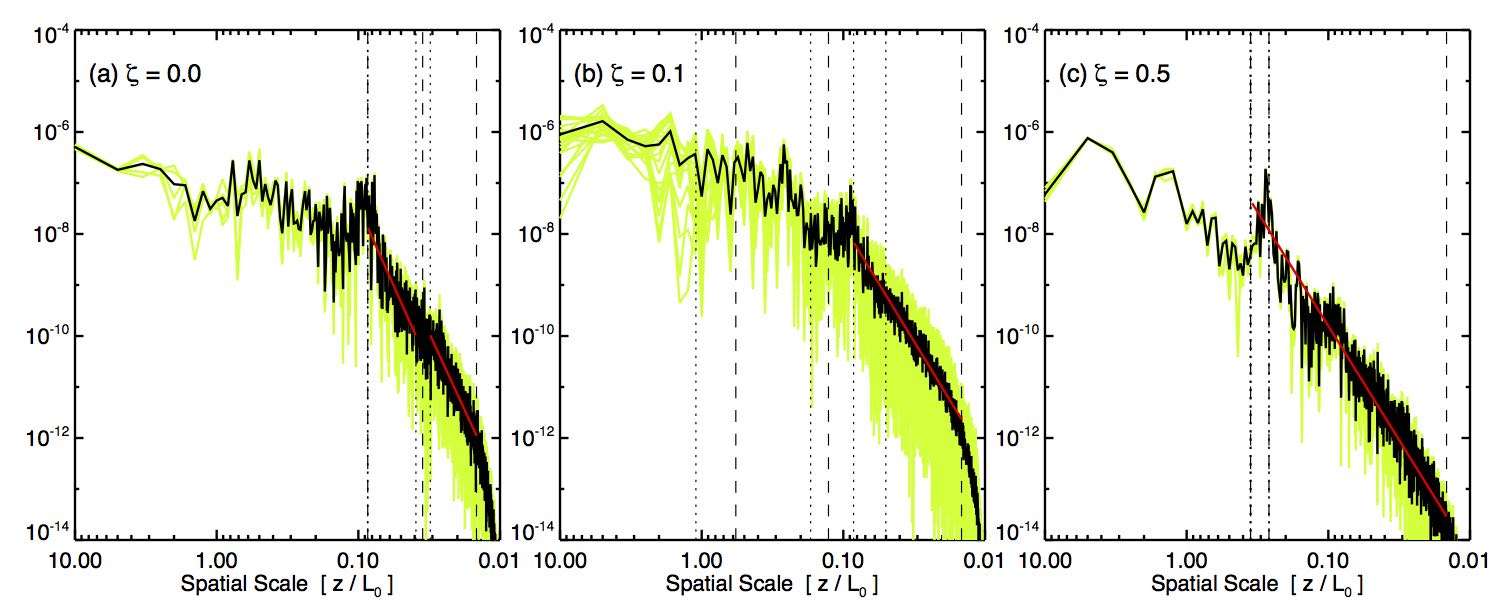}}
	\centerline{\includegraphics[width=1.0\textwidth]{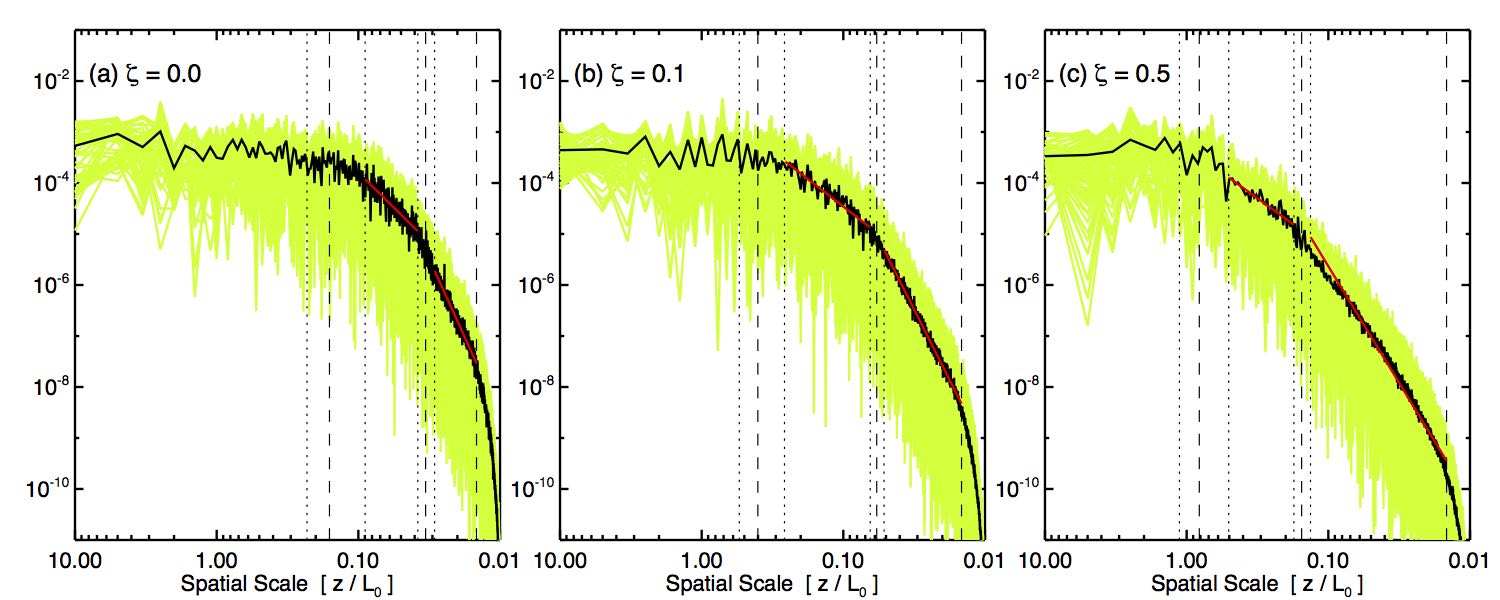}}
	\caption{Density fluctuation Fourier power spectra: (a) $\zeta = 0.0$; (b) $\zeta = 0.1$; 
	and (c) $\zeta = 0.5$. The green are every individual FFT power spectra from 1D 
	$\rho( z )/\rho_0$ along the current sheet. The solid black is the average FFT power 
	spectra. The dashed and dotted vertical lines indicate the minimum structure formation 
	coherency scale ($\xi_{I}$) and maximum inertial range scale ($\xi_{A}$) with respective $1\sigma$ uncertainties. 
	The top row exhibits spectra from the linear phase at times $t / \tau_A = \left\{ 49.08, 52.35, 68.46 \right\}$, respectively; 
	the bottom row exhibits spectra during the non-linear phase at time $t = 109.229 \ \tau_A$.
	}
\label{NLCohScls}
\end{figure*}

Figure~\ref{NLCohScls} plot representative one dimensional Fourier power spectra of the 
density enhancement fluctuations within the current sheet for each guide field case 
$\zeta = \left\{ 0.0, 0.1, 0.5 \right\}$ during the linear and non-linear phases, respectively, as a function of 
longitudinal ($z$-direction) spatial scale $\xi$. Since a relatively strong guide field suppresses non-linear 
onset, the weak $\zeta = 0.0$ case has turned non-linear before the current sheet is identifiable in the 
relatively strong $\zeta = 0.5$ case; hence the top row linear phase $\zeta = \left\{ 0.0, 0.1, 0.5 \right\}$ guide field cases are plotted at times $t / \tau_A = \left\{ 49.08, 52.35, 68.46 \right\}$, respectively. Figure~\ref{NLCohScls} bottom row however, shows all guide field cases well-into the non-linear phases, all at the fixed time $t = 109.229 \ \tau_A$. As in previous plots, the green curves 
are the entire set of Fourier spectra for each 1D-data along the $z$-direction. The thick 
black line is the mean spectra. The dashed and dotted vertical lines indicate the 
minimum structure formation coherency scale ($\xi_{I}$) and maximum inertial range scale ($\xi_{A}$) with 
respective $1\sigma$ uncertainties.

For each simulation, we seek a power-law scaling relation for the distinct characteristic canonical turbulence ranges identified in the Fourier power spectra. The red slopes of Figure~\ref{NLCohScls} illustrate these power-law fits over their respective spectral ranges for, respectively, the energy continuation range $\xi^{-\alpha \pm \sigma_{\alpha}}$ and inertial $\xi^{-\beta \pm \sigma_{\beta}}$ range for all guide field strengths. The linear phase spectra (top row) show two characteristic regions: an energy injection region dominating the majority of large spatial scales, and an inertial range beginning at typical scales of order $0.1 - 0.3 \ z/L_{0}$. In contrast, the non-linear phase spectra (bottom row) of all cases shown exhibit three characteristic regions in their respective nonlinear phase: energy injection, energy continuation, and inertial range.

\begin{figure}
	\centerline{\includegraphics[width=0.65\textwidth]{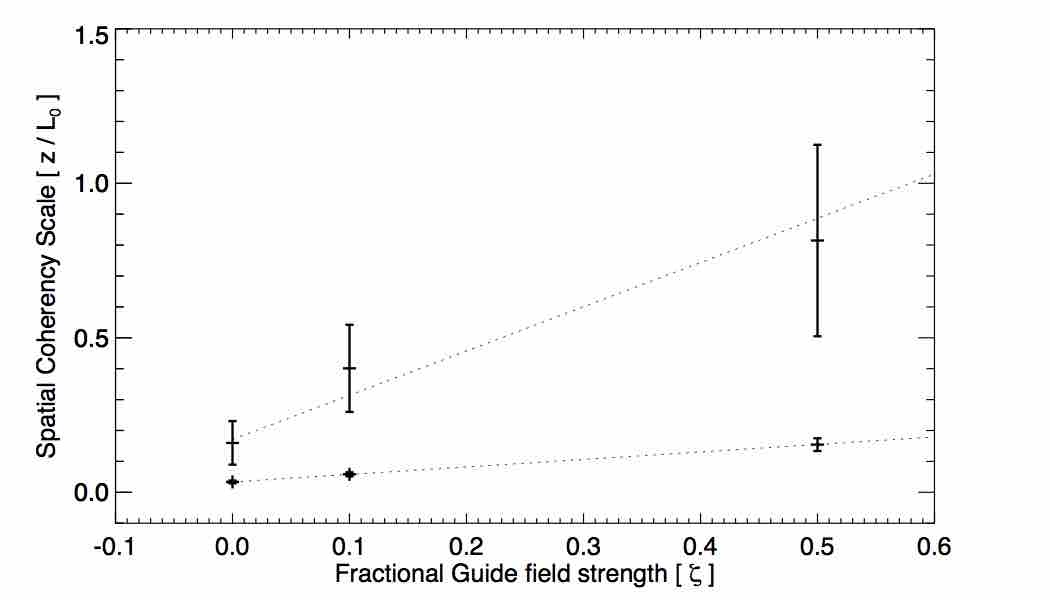}}
	\caption{Longitudinal coherency scale dependence on normalized guide field strength $\zeta$. 
	The three upper points show the minimum structure formation coherency scale $\xi_{I}$ and the 
	three lower points show the maximum inertial range scale $\xi_{A}$.}
\label{CohScls2}
\end{figure}

In Figure~\ref{CohScls2} we plot the two identified longitudinal coherency scales' 
$\xi_{I}$ (upper data) and $\xi_{A}$ (lower data), as a function of the normalized guide 
field strength $\zeta$. To within the calculated $\pm 1 \sigma$ uncertainties, the increase 
in longitudinal spatial scale increases reasonably linearly with increasing (normalized) 
guide field strength.

Table~\ref{Tbl1} lists the time-averaged structure coherency scales, $\xi_{I}$, $\xi_{A}$, 
with their respective $\pm 1 \sigma$ uncertainties, as well as the time-averaged power-law 
exponents, $\alpha$, $\beta$, with their respective $\pm 1 \sigma$ uncertainties, obtained 
from all output data files (every 0.5 $\tau_A$) in each simulation for $109.229 \le t/\tau_A \le 120$. 
It is clear from Figures \ref{NLCohScls} and \ref{CohScls2}, as well as Table~\ref{Tbl1}, that all 
of our structure coherency scales increase with guide field strength, as expected. Furthermore, 
the energy continuation exponents are constant and independent guide field 
strengths to within their respective $\pm 1 \sigma$ uncertainties, whereas there is a statistically 
significant decrease in the inertial range exponent with increasing guide field. The increased width of the inertial range with increased guide field strength, as well as the softer the power-law slope, represents the continued development of a turbulence-like distribution of density fluctuation scale sizes in the plane of the current sheet. As the characteristic ``energy injection'' length scale of magnetic island plasmoids increase with the guide field strength, there is more opportunity for the kinking, break up, and other nonlinear interactions to cascade this fluctuation``energy'' through a broader range of scale sizes. 

\begin{table}
\begin{center}
\begin{tabular}{| c | c | c | c | c | c |}
\hline
Guide Field 	& Min. Formation	& Eng. Continuation Rng.	& Max. Inertial	& Inertial Rng.	& Dissipation \\[-0.1in]
Strength		& Scale			& Exponent			& Scale		& Exponent	& Scale \\[-0.1in]
$\zeta$ 		& $\xi_{I} \pm \sigma_{I}$ & $\alpha \pm \sigma_{\alpha}$ & $\xi_{A} \pm \sigma_{A}$ & $\beta \pm \sigma_{\beta}$ & $\xi_{D}$ \\
\hline\hline
\ 0.0 \ & \ 0.160 $\pm$ 0.070 \ & \ 2.875 $\pm$ 0.930 \ & \ 0.032 $\pm$ 0.004 \ & \ 6.161 $\pm$ 0.994 \ & \ 0.015 \ \\
\ 0.1 \ & \ 0.401 $\pm$ 0.141 \ & \ 2.225 $\pm$ 0.615 \ & \ 0.056 $\pm$ 0.006 \ & \ 5.541 $\pm$ 0.347 \ & \ 0.015 \ \\
\ 0.5 \ & \ 0.815 $\pm$ 0.310 \ & \ 2.004 $\pm$ 1.330 \ & \ 0.151 $\pm$ 0.020 \ & \ 4.582 $\pm$ 0.166 \ & \ 0.015 \ \\
\hline
\end{tabular}
\end{center}
\caption{Power-law fit exponents for each of the spatial scale ranges in the density Fourier power spectra described in Section~\ref{SSS:FP} for simulations $\zeta=\{ 0.0, 0.1, 0.5 \}$. The power-law exponent $\alpha$ corresponds to the continuation range, $\xi_{I} \ge z/L_0 \ge \xi_{A}$, defined between the maximum wavelet IPPS scale ($\xi_{I}$) and the autocorrelation spatial lag ($\xi_{A}$). The power-law exponent $\beta$ corresponds to the inertial range, $\xi_{A} \ge z/L_0 \ge \xi_{D}$, defined between $\xi_{A}$ and the dissipation scale $\xi_{D}$. See text for details.\\} \label{Tbl1}
\end{table}

\section{Reconnection-Driven Solar Wind Outflow from Pseudostreamers} \label{S:Discussion}

Our simulation results have the following implications for reconnection-driven solar wind outflow from pseudostreamers.

First, magnetic reconnection generates highly structured density fluctuations corresponding to 3D magnetic islands even under relatively gentle (i.e. quasi-steady state) reconnection inflow conditions. These plasmoids propagate through the current sheet and either become part of the pseudostreamer's closed flux system or are on magnetic field lines that have interchange-reconnected and are part of the structured, intermittent reconnection outflow along the pseudostreamer external spine fan corresponding to the Separatrix-Web arc \citep{Antiochos2011,Higginson2017b}. Our results confirm the smooth interchange reconnection pseudostreamer outflow scenario of \citet{Masson2014b} and show with a sufficiently resolved current sheet, the plasmoid instability during reconnection provides a potential explanation for some of the variability observed in pseudostreamer wind \citep{Crooker2012a, Crooker2014, Owens2014}.

Second, the characteristic coherency scales of our reconnection-generated density fluctuations in the plane of the current sheet are strongly a function of guide field strength. In other words, with no guide field ($\zeta=0$), the density fluctuations are essentially isotropic with a scale of $\sim$400~km whereas in the weak ($\zeta=0.1$) and moderate ($\zeta=0.5$) guide field cases they become increasingly elongated in the guide field direction with scales of $\sim$600~km and $\sim$1500~km, respectively. Given the transverse scales remain constant at $\sim$400~km, this corresponds to an increasing guide-field aligned density fluctuation axial ratio of 1.0, 1.5, and 3.8. Radio, white light, and EUV remote sensing observations show significant field-aligned and highly-collimated density structures/striations \citep[e.g.][]{Woo2003,  Druckmuller2014, Raymond2014}. Interplanetary scintillation observations by \citet{Grall1997} showed relatively high axial ratios for electron density fluctuations $\delta N_e$ of $\gtrsim 10$ close to the sun ($\lesssim 5 \ R_\odot$) that decreased rapidly to $\lesssim 3$ for distances $> 5 \ R_\odot$. 
Our idealized simulations' density fluctuation axial ratios are consistent with the lower end of the observed ranges and we are looking forward to investigating the propagation and evolution of these signatures into a solar wind outflow \citep[e.g.][]{Karpen2017, Uritskyetal2017}.   
 
Third, the coherency of the magnetic structure of the 3D magnetic island flux ropes in the guide field direction is also directly related to the guide field strength. The representative magnetic field lines in the Figure~\ref{FLICx} visualizations show the increasing flux rope sizes in the plane of the current sheet. \citet{Wyper2016b} described the formation and evolution of 3D magnetic island flux ropes during the reconnection associated with coronal jets and essentially the same picture holds for our extended pseudostreamer arcade as well. Reconnection forms a localized region of highly twisted field in the 3D magnetic island but the field lines that make up this newly-reconnected open flux tube are relatively untwisted far from the current sheet. Therefore, the twist itself will propagate away as torsional Alfv\'{e}nic waves from the reconnection site in both directions in an attempt to spread out along the field lines as much as possible (see $\S$5.2 and Figure~9 of \citealt{Wyper2016b}). These Alfv\'{e}nic fluctuations propagate both down the flux tube toward the lower boundary and along the direction of the flux tube that opens up into the solar wind. Thus, depending on the guide field strength during the interchange reconnection process, in addition to highly-structured density fluctuation outflow, one may also see periods of highly-structured magnetic field fluctuations or Alfv\'{e}nic field rotations in pseudostreamer wind \citep[e.g., see][]{Higginson2017c}.

\section{Summary and Conclusions} \label{S:Conc}

We have extended the analysis of \citetalias{Edmondson2010b} by analyzing a new set of 3D MHD simulations with a systematically increasing guide field component. We quantified both the longitudinal and transverse coherency scales of the mass density fluctuations generated during the nonlinear phase of plasmoid-unstable reconnection in a model solar corona and have shown that the increasing guide field strength has two main effects on our system. First, the onset of magnetic reconnection and transition to the nonlinear phase of the plasmoid instability is delayed with increasing guide field component. Second, the longitudinal coherency of the reconnection-generated 3D magnetic island flux ropes significantly increases with guide field strength. 

In this set of simulations, the delay in nonlinear plasmoid-unstable onset is due to the numerical resistivity which depends on the local gradients in the magnetic fields and flows. The transition from strictly anti-parallel reconnection (no guide field) to component reconnection (a strong guide field) corresponds to a smoothing out of the total field gradient across the current sheet. This smoothing has the effect of reducing the effectiveness of the numerical diffusion. This result is well known from previous reconnection studies and we have demonstrated its applicability to systems with fully consistent, 3D current sheet formation in response to the imposed separation of the spine lines in the Syrovatskii scenario.

We have established the correspondence between the plasmoid density enhancement fluctuations in the plane of the current sheet with the 3D magnetic island flux rope structures, and hence the coherency scales of the density enhancements as a proxy for the full flux rope structures. Using standard spectral and correlation methods, we demonstrated the longitudinal structure coherency increases with guide field, whereas the transverse structure coherency remains independent of the guid field strength.

We interpret the Fourier power spectra of the plasmoid density fluctuations within the current sheet with respect to canonical turbulence, defining the corresponding energy injection range (of coherent structure formation), the energy continuation range transition, and an inertial range. We quantified how the guide field impacts the spatial scale boundaries of these ranges and the spectral slopes in each of these regions.

Finally, we have discussed the implications of our simulation results for properties of interchange-reconnection generated 
solar wind outflow in the vicinity of coronal pseudostreamers.

\acknowledgments

J.K.E. and B.J.L. acknowledge support from NASA NNX15AJ66G, NNX15AB69G, and NSF AGS-1622495. The numerical simulations were made possible with an allocation of NASA High-End Computing resources and were carried out on the Discover cluster at the NASA Center for Climate Simulation at Goddard Space Flight Center.








\appendix

\section{Improved Estimate of the Numerical Resistivity} \label{SB:NumRes}

In this appendix, we derive an improved estimate of the numerical resistivity. This improved estimate is based on a simple energy density transport through the dissipative region. In other words, we balance the convective transport of the kinetic and magnetic energy densities against the Ohmic dissipation within the current sheet,
\begin{equation} \label{TransportEqn}
\oint_{A} \ \left( \ \frac{B^{2}}{8 \pi} + \frac{\rho V^{2}}{2} \ \right) \bm{V} \cdot {\rm d}\bm{S} = \int_{V} \ \eta \ J^{2} \ {\rm d}V.
\end{equation}

We consider the dissipation region to be a volume with dimensions: cross-sectional thickness $\delta$ and cross-sectional length $L$, and perpendicular depth $D$. Relative to the global Cartesian coordinates used in this paper, the cross-sectional plane is spanned by the $x$-$y$ directions, and the perpendicular depth is identified with the $z$-direction. Hence, we may write equation (\ref{TransportEqn}) as
\begin{equation} \label{TransportEqnSimp}
\left( \ \frac{B^{2}}{8 \pi} + \frac{\rho V^{2}}{2} \ \right)_{in} V_{in} \ L \sim \left( \ \frac{B^{2}}{8 \pi} + \frac{\rho V^{2}}{2} \ \right)_{out} V_{out} \ \delta + \eta \ J^{2} \ L \delta \\
\end{equation}

For simplicity, and without loss of generality, we assume the in-states, and out-states, are equivalent on the respective right/left boundaries of the current sheet ($V_{in}^{(+)} = V_{in}^{(-)}$, $V_{out}^{(+)} = V_{out}^{(-)}$, $B_{in}^{(+)} = B_{in}^{(-)}$, and $B_{out}^{(+)} = B_{out}^{(-)}$).

We estimate $J^{2}$ through the dissipation region's cross sectional area from Ampere's Law (in cgs units),
\begin{equation} \label{Ampere}
\frac{4 \pi}{c} \ \int_{A} \ \bm{J} \cdot {\rm d}\bm{S} = \oint_{C} \bm{B} \cdot {\rm d}\boldsymbol{\ell} \\
\end{equation}

\noindent Applied to the dissipation volume
\begin{equation} \label{AmpereSimple}
\frac{4 \pi}{c} \ J \ L \ \delta \sim 2 \ B_{in} \ L + 2 \ B_{out} \ \delta \\
\end{equation}

\noindent Hence, we estimate $J^{2}$ in terms of $B_{in}$, $B_{out}$, and the cross sectional geometry
\begin{equation} \label{jSqrd}
J^{2} \sim \left( \ \frac{c}{4 \pi} \ \right)^{2} \left( \ \frac{2 B_{in}}{\delta} \ \right)^{2} \left( \ 1 + \frac{B_{out}}{B_{in}} \ \frac{\delta}{L} \ \right)^{2} \\
\end{equation}

Substituting (\ref{jSqrd}) into (\ref{TransportEqnSimp}), we estimate the numerical resistivity scales as
\begin{equation} \label{NumResistScale}
\eta \sim \frac{4 \pi}{c^{2}} \ \frac{V_{in} \ \delta}{4} \left( \ 1 + \frac{B_{out}}{B_{in}} \ \frac{\delta}{L} \ \right)^{-2} \left[ \ \left( \ 1 + \frac{V_{in}^{2}}{V_{Ain}^{2}} \ \right) - \frac{V_{out}}{V_{in}} \left( \ \frac{B_{out}^{2}}{B_{in}^{2}} + \frac{\rho_{out}}{\rho_{in}} \ \frac{V_{out}^{2}}{V_{Ain}^{2}} \ \right) \frac{\delta}{L} \ \right] \\
\end{equation}

\noindent where $V_{Ain} = B_{in} / \sqrt{4 \pi \rho_{in}}$ is the (average) Alfv\'{e}n speed in the in-flow region. 

Hence, modulo the constant factor $4 \pi / c^2$, we estimate the numerical resistivity as
\begin{equation} \label{NumResistScale1stOrder}
\eta \sim V_{in} \ \delta \ \frac{\left( \ \chi + Z \ \right)}{4}. \\
\end{equation}

\noindent We define, to first order in $\delta / L$, the dimensionless factors for the advection of magnetic energy density $\chi$ and compressibility $Z$, respectively,
\begin{equation}
\chi = 1 - \left( \ 2 \ \frac{B_{out}}{B_{in}} + \frac{V_{out}}{V_{in}} \ \frac{\mathcal{ME}_{out}}{\mathcal{ME}_{in}} \ \right) \frac{\delta}{L} \\
\end{equation}
\begin{equation}
Z = M_{A}^{2} \left[ \ 1 - \left( \ 2 \ \frac{B_{out}}{B_{in}} + \frac{V_{out}}{V_{in}} \ \frac{\mathcal{KE}_{out}}{\mathcal{KE}_{in}} \ \right) \frac{\delta}{L} \ \right] \\
\end{equation}

\noindent Where $\mathcal{ME} = B^{2} / (8 \pi)$ is the magnetic energy density advected with the in/out-flows, $\mathcal{KE} = \rho \ V^{2} / 2$ is the kinetic energy density advected with the in/out-flows, and $M_{A} = V_{in} / V_{Ain}$ is the Alfv\'{e}nic Mach number of the in-flow.

The results in Figure~\ref{CSprop}, including the current sheet aspect ratio, $\eta$, and $S$ for each of the simulation cases $\zeta = \{ 0, 0.1, 0.5, 1.0 \}$ are calculated using the current sheet length estimate $L_{cs}(t)$ and time-dependent, current sheet-averaged quantities $\langle V_{in} \rangle$, $\langle V_{out} \rangle$, $\langle B^2_{in} \rangle$, $\langle B^2_{out} \rangle$, and $\langle V_{Ain} \rangle$ described in the next section.

\section{Constructing Average Quantities in Current Sheet Coordinates} \label{SA:CSFit}

We use a slightly simpler and more automated version of the procedure outlined in the Appendices of \citet{Lynch2016a} to define a current sheet-aligned coordinate transform to construct the average current sheet quantities required for the estimates of the Lundquist number $S(t)$, the numerical resistivity estimate $\eta(t)$, and to analyze the mass density fluctuations and their evolution in the plane of the current sheet in Section~\ref{S:Results}.

Using the spatial position of grid cells in the $x$-$y$ plane at $z=0$ at each simulation output time $t_k$, we perform a linear least-squares fit $f_{k}(x) = c_0 + c_1 x$, to the quantity $|J(x_i,y_j)| / |B(x_i,y_j)| \ge \Gamma$ where the parameter $\Gamma$ varies with guide field strength ($\Gamma \ge 100$ for $\zeta = 0$; $\Gamma \ge 31.6$ for $\zeta = 0.1$; $\Gamma \ge 100$ for $\zeta = 0.5$, $\Gamma \ge 15$ for $\zeta = 1.0$). We assign these spatial positions weights based on the current density magnitude $\sigma_{J} = \left( J(x_i,y_j) / J_{max} \right)^4$. 

Furthermore, we define the minimum and maximum $x$-values of the spatial positions meeting a less-restrictive $\Gamma$ criteria ($\Gamma \ge 31.6$ for $\zeta = 0$; $\Gamma \ge 5.0$ for $\zeta = 0.1$; $\Gamma \ge 40.0$ for $\zeta = 0.5$, $\Gamma \ge 2.0$ for $\zeta = 1.0$) as $\left[ x_0, x_1\right]$ and construct their corresponding $y$-values from the current sheet fit, $\left[ y_0, y_1\right] = \left[ f_k(x_0), f_k(x_1) \right]$. The current sheet length is then simply calculated as the distance between the current sheet endpoints, $2L_{cs} = \sqrt{(x_1-x_0)^2 + (y_1-y_0)^2}$. 

\begin{figure}
	\centerline{\includegraphics[width=0.9\textwidth]{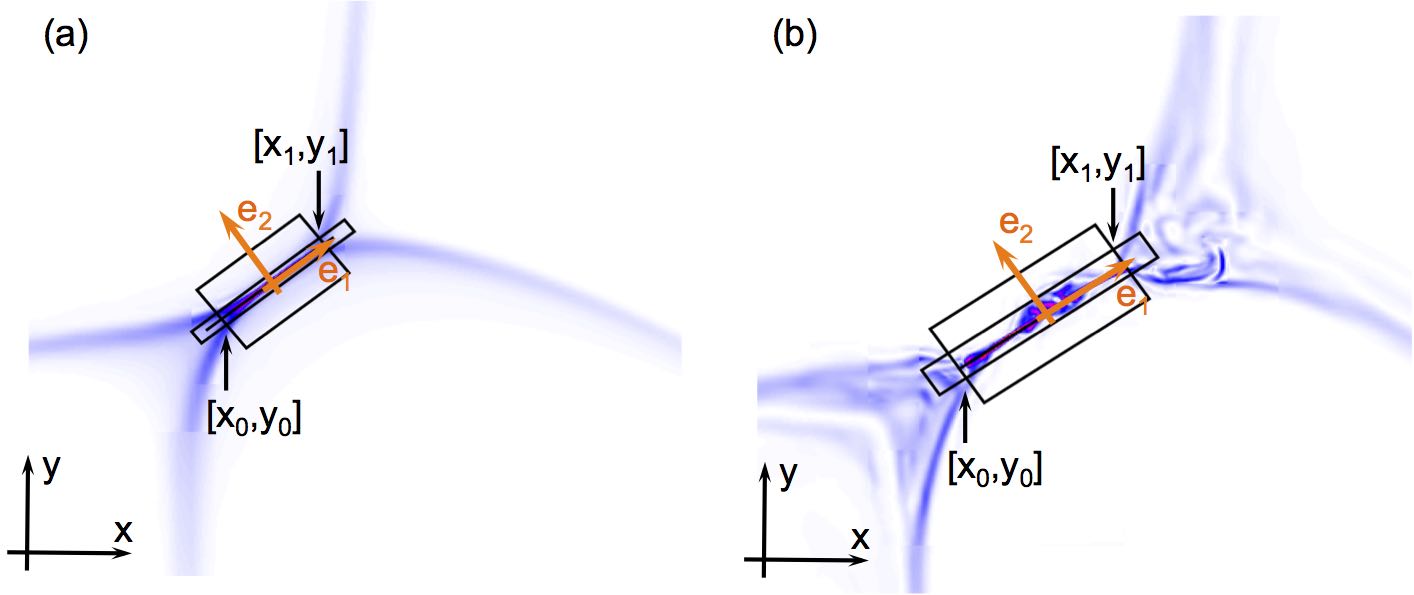}}
	\caption{Current sheet identification: panel (a) automated procedure during laminar growth; panel (b) frame-by-frame user-based visual inspection procedure after non-linear onset. The current sheet coordinate direction vectors $\lbrace \ \hat{\boldsymbol{e}}_1, \hat{\boldsymbol{e}}_2 \ \rbrace$ are shown in orange. The current sheet is identified by the Y-point termini and the width of the largest plasmoid.}
\label{CSIden}
\end{figure}

We note, the above automated identification procedure worked reasonably well to fit the laminar current sheet formation stage, prior to non-linear onset (see Figure \ref{CSIden} panel (a)). However, in the non-linear stage, the current sheet dynamics of plasmoid formation/ejection and re-formation of the current sheet required frame-by-frame user-based visual inspection and modification of the $\left[x_0,x_1\right]$ and $\left[ y_0, y_1 \right]$ values. This simple, albeit tedious identification procedure was informed by identifying the local current sheet Y-point termini for the current sheet length, and including the largest instantaneous plasmoid for the width (see Figure \ref{CSIden} panel (b)).

The linear fit to the current sheet position is also used to define the current sheet coordinates $\lbrace \ \hat{\boldsymbol{e}}_1, \hat{\boldsymbol{e}}_2 \ \rbrace$, given by
\begin{equation}
\left[ \begin{array}{c} \hat{\bm{e}}_1 \\ \hat{\bm{e}}_2 \end{array} \right] = 
\frac{1}{\sqrt{1 + c_1^2}}
\left[ \begin{array}{cc} 1 &  c_1 \\  -c_1  & 1 \end{array} \right]
\left[ \begin{array}{c} \hat{\bm{x}} \\ \hat{\bm{y}} \end{array} \right] ,
\end{equation}  
where $\hat{\boldsymbol{e}}_1$ is parallel to the linear current sheet fit, $f_k$, and perpendicular $\hat{\boldsymbol{e}}_2$ (see Figure \ref{CSIden}). The current sheet reconnection inflow and outflow velocity projections are now readily obtained via $V_{in} = \bm{V} \cdot \hat{\boldsymbol{e}}_2$ and $V_{out} = \bm{V} \cdot \hat{\boldsymbol{e}}_1$. 

In order to construct the current sheet-averaged quantities $\langle V_{in} \rangle$, $\langle V_{out} \rangle$, $\langle B^2_{in} \rangle$, and $\langle B^2_{out} \rangle$, we define the inflow rectangular regions as centered on the $f_k(x)$ line extending $\pm10\delta$ in $\hat{\boldsymbol{e}}_2$, and the outflow rectangular regions extending $10\delta$ in $\hat{\boldsymbol{e}}_1$ from each end of the current sheet (see Figure \ref{CSIden}). 
%

\bibliographystyle{aasjournal}
\bibliography{master_20170221} 




\end{document}